\documentclass[aps,prc,twocolumn,groupedaddress,showpacs]{revtex4}
\usepackage{graphicx}
\usepackage{amsmath}

\def\eqref#1{Eq.~(\ref{eq:#1})}
\def\eqlab#1{\label{eq:#1}}
\def\figref#1{Fig.~(\ref{fig:#1})}
\def\figlab#1{\label{fig:#1}}
\def\tabref#1{Tab.~(\ref{tab:#1})}
\def\tablab#1{\label{tab:#1}}
\def\seclab#1{\label{sect:#1}}
\def\secref#1{Section~\ref{sect:#1}}

\begin{document}


\title{Constraining the density dependence of the symmetry energy\\ using the multiplicity and average $p_T$ ratios of charged pions}


\author{M.D. Cozma}
\email{dan.cozma@theory.nipne.ro}
\affiliation{Department of Theoretical Physics, IFIN-HH,\\ Reactorului 30, 077125 M\v{a}gurele/Bucharest, Romania}


\date{\today}

\begin{abstract}
The charged pion multiplicity ratio in intermediate energy heavy-ion collisions, 
a probe of the density dependence of symmetry energy
above the saturation point, has been proven in a previous study to be extremely sensitive to 
the strength of the isovector $\Delta$(1232) potential in nuclear matter. As there is no knowledge, 
either from theory or experiment, about the magnitude of this quantity, the extraction 
of constraints on the slope of the symmetry energy at saturation by using exclusively the mentioned observable
is hindered at present. It is shown that, by including the ratio
of average $p_T$ of charged pions $\langle p_T^{(\pi^+)}\rangle/\langle p_T^{(\pi^-)}\rangle$
in the list of fitted observables, the noted problem can be circumvented. 
A realistic description of this observable requires accounting 
for the interaction of pions with the dense nuclear matter environment by the incorporation of the so called 
$S$-wave and $P$-wave pion optical potentials. 
This is performed within the framework of a quantum molecular dynamics transport model that enforces the conservation of the total
energy of the system. It is shown that constraints 
on the slope of the symmetry energy at saturation density and the strength of the $\Delta$(1232) potential
can be simultaneously extracted. A symmetry energy with a value of the slope parameter $L>$ 50 MeV is
favored, at 1$\sigma$ confidence level, from a comparison with published FOPI experimental data. A precise constraint will
require experimental data more accurate than presently available, particularly for the charged pion multiplicity ratio, and
better knowledge of the density and momentum dependence of the pion potential for the whole range of these two variables
probed in intermediate energy heavy-ion collisions.

\end{abstract}


\maketitle

\section{Introduction}
Pions produced in intermediate energy heavy-ion collisions have been shown to provide promising means
to study the isovector part of the equation of state (asy-EoS) of nuclear matter, commonly known as the
symmetry energy (SE). The multiplicity ratio of charged pions (PMR) has been proven to be sensitive to the density
dependence of SE~\cite{Li:2004cq}, particularly to the density range of half to twice saturation density ($\rho_0$), with a maximum 
in sensitivity around 1.25$\rho_0$~\cite{Liu:2014psa}, while uncertainties in the isoscalar part of the equation of state are suppressed.
This makes it suitable for extracting constraints for the value of the slope of the symmetry energy at saturation once its magnitude
at saturation or at other particular density value is known from other sources (e.g., nuclear structure studies
~\cite{Brown:2013mga,Zhang:2013wna}). Higher order terms (e.g., curvature term and the associated $K_{sym}$ parameter), while
potentially important for the extrapolation of the symmetry energy to densities of interest for astrophysics studies,
are customarily assumed to have a small effect and consequently simpler, one free parameter, parametrizations are adopted in 
heavy-ion transport calculations (e.g., the Gogny inspired MDI interaction~\cite{Das:2002fr}). The impact of SE on PMR has 
been shown to grow larger as the energy of the incident beam is decreased.

Attempts to constrain the slope of the SE at saturation by making use of various transport models
and the experimentally measured value for the PMR in central $^{197}$Au+$^{197}$Au at an impact energy
of 400 MeV/nucleon have resulted in a confusing picture: constraints on the high density dependence of the SE ranging
from a very soft to a stiff one have been extracted~\cite{Xiao:2008vm,Xie:2013np,Feng:2009am}, 
or even no sensitivity on the slope parameter has been reported~\cite{Hong:2013yva}. Additionally,
most models have led to a contradiction between the $\pi^-$/$\pi^+$ multiplicity ratio and neutron/proton elliptic
flow ratio extracted constraints for the SE stiffness. Efforts to find a solution to this problem by studying the impact of  in-medium 
modifications of the pion-nucleon interaction~\cite{Xu:2013aza}, the
kinetic part of the SE term~\cite{Li:2014vua}, the neutron skin thickness~\cite{Wei:2013sfa}, or particle production threshold shifts 
due to the inclusion of self-energy contributions~\cite{Ferini:2006je,Song:2015hua} on the PMR value have proven, from
a quantitative point of view, largely unsuccessful, but some interesting findings were nevertheless reported.

The impact of including the self-energy contributions in the constraint of energy conservation that appears
in the collision term of the transport equations, and thus implicitly
modifying particle production thresholds, has been explored in Refs.~\cite{Ferini:2006je,Song:2015hua}. Such an approach
leads to a manifest implementation of energy conservation at local level; $\it i.e.$, only the total energy of the particles
involved in a binary reaction is conserved. It has lead to the interesting result that a stiffer asy-EoS leads to a slightly 
larger PMR than a soft choice would, which is opposite to the result obtained when the self-energy contributions to the energy
conservation constraint are neglected. The effect was however found not to be quantitatively large enough to allow the extraction,
from a comparison with experimental data, of the value for the slope $L$ of the SE at saturation.

The next step was taken in Ref.~\cite{Cozma:2014yna}, where a transport model which enforces the conservation of the total energy
of the entire system during heavy-ion reactions has been developed. A restriction of the model to the so-called local energy 
conservation scenario, which resembles the models of Refs.~\cite{Ferini:2006je,Song:2015hua} closest (up to relativistic corrections
of the dynamics) due to the relationship between self-energies and effective potentials
(the latter being related to the real part of the former),
 has confirmed the results of those studies. The requirement of global total energy
conservation was reported to have an important impact on pion multiplicities, particularly $\pi^-$, preserving the sensitivity of
the PMR to the SE stiffness, but enhancing the effect of a higher multiplicity ratio for a stiffer asy-EoS reported in 
Ref.~\cite{Song:2015hua}. However, a large dependence of the PMR's magnitude on the strength of the 
isovector part of the $\Delta$(1232) potential was evidenced, which, in view of the lack of 
information on this quantity, rendered this observable unsuitable for constraining the density dependence of the SE. 
This is in contrast with conclusions regarding the impact of the $\Delta$(1232) potential on pionic observables reached in 
Refs.~\cite{Li:2015hfa,Guo:2015tra}. It is however not a conflict since in these lastly mentioned studies the threshold effects
generated by the conservation of the total energy have not been accounted for, the impact of the $\Delta$ (1232) baryon arising
only due to its motion in the mean field. Owing to the short lifetime of this resonance, the impact of its in-medium potential
on pionic spectra is modest for values of the impact energy for which experimental data are available. 
Another important conclusion of the  study in Ref.~\cite{Cozma:2014yna} was that for the standard choice for 
the strength of the isovector $\Delta$(1232) potential, equal to that of the nucleon, an almost perfect agreement 
between the pion and elliptic flow extracted SE constraints could be obtained. 

The present study extends the analysis performed in Ref.~\cite{Cozma:2014yna} to the average $p_T$ ratio of charged pions,
$\langle p_T^{(\pi^+)}\rangle/\langle p_T^{(\pi^-)}\rangle$ (PAPTR). It is shown that by using both observables, PMR and PAPTR, 
constraints on the stiffness of the SE can be extracted, independently of the strength of the isovector $\Delta$(1232) potential. 
Constraints on the latter are naturally a by-product of such a study. To achieve this goal, the model of Ref.~\cite{Cozma:2014yna} is further
improved by including the optical potential of pions in nuclear matter, both the so-called $S$- and $P$-wave
components~\cite{Kisslinger:1955zz,Ericson:1966fm}. All the relevant details of this development are presented in Sec. II. 
The impact of this quantity on multiplicities, multiplicity spectra, and average $p_T$ values 
of pions is studied in detail, and, where available, a comparison with experimental FOPI data
~\cite{Reisdorf:2006ie,Reisdorf:2010aa,Reisdorf:2015aa} is presented. Additionally, the impact
of poorer known model parameters is also investigated, followed by a presentation of the extracted constraints for SE (Sec. III). 
The article ends with a section devoted to summary and conclusions.

\section{The model}
\seclab{themodel}
\subsection{The Transport Model}
\seclab{transpmodel}
Heavy-ion collision dynamics is simulated using an upgraded version~\cite{Cozma:2014yna} of the T\"ubingen quantum
molecular dynamics Model (QMD) transport model~\cite{Khoa:1992zz,UmaMaheswari:1997ig} which provides a 
semiclassical framework for the description of such reactions and accounts for relevant quantum aspects such as 
stochastic scattering and Pauli blocking of nucleons.
It includes the production of all nucleonic resonances with masses below 2 GeV, in total 11 N$^*$ and 10 $\Delta$ resonances.
At energies of interest for this study pions are produced predominantly by the excitation of the $\Delta$(1232) isobar in
inelastic nucleon-nucleon collisions.

In QMD-type transport models, the total wave function of the ensemble of nucleons is taken to be the product of individual
nucleon wave functions which are each represented by a Gaussian wave packet of finite spread in phase space.
To make the transition to a semi-classical picture a formulation of quantum mechanics that is obtained by applying the Weyl
transformation to the standard Schr\"odinger one is employed. The Wigner distribution, which is defined as
the Weyl transform of the statistical density operator, is introduced. It represents the quantum analogue of classical phase 
space densities with the exception that it can take both positive and negative values. With its help, it can be shown
that the expectation values of the position and momentum operators satisfy the classical Hamiltonian equations of motion
~\cite{deGroot:1972aa,Hartnack:1997ez} which can be factorized to each particle given the Ansatz made for the total wave function of the system,
\begin{eqnarray}
\frac{d\vec{r}_i}{dt}=\frac{\partial \langle U_i \rangle}{\partial \vec{p}_i}+\frac{\vec{p}_i}{m},\qquad
\frac{d\vec{p}_i}{dt}=-\frac{\partial \langle U_i \rangle}{\partial \vec{r}_i}\,.
\end{eqnarray}
Here, the average of the potential operator is understood to be taken over the entire phase-space and weighted by the Wigner distribution
of particle $i$. The potential operator $U_i$ is in this case the sum of the Coulomb and strong interaction potential operators.
In all kinematic equations the relativistic relation between mass, energy and momentum is used.

The Gogny-inspired parametrization of the equation of state of
nuclear matter~\cite{Das:2002fr} has been selected to describe the mean-field experienced by a nucleon at
finite density. It leads to a mean-field nucleon potential,
\begin{widetext}
\begin{eqnarray}
\eqlab{eqsympot}
 U(\rho,\beta,p,\tau,x)&=&A_u(x)\frac{\rho_{\tau'}}{\rho_0}+A_l(x)\frac{\rho_{\tau}}{\rho_0}
+B\,\Big(\frac{\rho}{\rho_0}\Big)^\sigma(1-x\beta^2)
-8\tau x\frac{B}{\sigma+1}\frac{\rho^{\sigma-1}}{\rho_0^\sigma}\beta\rho_{\tau'} \\
&&+\frac{2C_{\tau \tau}}{\rho_0}\int d^{\!\:3} \vec{p}\!\;'\, \frac{f_\tau(\vec{r},\vec{p}\!\;')}{1+(\vec{p}-\vec{p}\!\;')^2/\Lambda^2} 
+\frac{2C_{\tau \tau'}}{\rho_0}\int d^{\!\:3} \vec{p}\!\;'\, \frac{f_{\tau'}(\vec{r},\vec{p}\!\;')}{1+(\vec{p}-\vec{p}\!\;')^2/\Lambda^2} \nonumber ,
\end{eqnarray}
\end{widetext}
that displays besides density ($\rho$) and isospin asymmetry ($\beta$) also a momentum ($p$) dependence in both 
the isoscalar and isovector components. The label $\tau$ designates the isospin component of the nucleon or resonance
while the parameter $x$ has been introduced to allow for an adjustment of the symmetry energy stiffness.
The isovector part of the Gogny interaction is reproduced by setting $x$=1.
Negative and positive values of this parameter correspond to a stiff and a soft density dependence, respectively. 
The values of the $C_{\tau \tau}$, $C_{\tau \tau'}$ and $\Lambda$ parameters are determined by optimally reproducing 
the momentum dependent part of the Gogny interaction~\cite{Das:2002fr}. This results in an effective isoscalar nucleon
mass of 0.7$m_N$ and a neutron-proton effective mass splitting of approximately 0.4$\beta$ at saturation density. The latter is in reasonable
agreement with the average of values put forward by presently undisputed studies which have aimed at determining it from 
experimental data~\cite{Xu:2010fh,Li:2014qta,Li:2013ola,Zhang:2015qdp}. The remaining parameters are determined from
the location of the saturation point ($\rho_0$), binding energy at saturation, magnitude of the symmetry energy at saturation ($S_0$=30.6 MeV) and value of the 
compressibility modulus ($K$=245 MeV). To be complete, the determined values of all parameters appearing in the
expression of the effective potential in~\eqref{eqsympot} read
\begin{align}
&\phantom{aaaaaaaaa}\Lambda=0.2630 \nonumber\\
&C_{\tau\tau}=-0.0117,\quad C_{\tau\tau'}=-0.1034 \nonumber\\
&\phantom{aaa}B=0.06844,\quad\sigma=1.57065\\
&\phantom{aaa}A_u(x)=-0.05807-\frac{2xB}{\sigma+1}\nonumber\\
&\phantom{aaa}A_l(x)=-0.08266+\frac{2xB}{\sigma+1}. \nonumber
\end{align}
The parameter $\sigma$ is dimensionless, the rest being expressed in units of GeV. The first three parameters take the
same values as in Ref.~\cite{Das:2002fr} while the others are different due to the chosen magnitude for the
compressibility modulus. Values for the slope ($L$) and curvature
($K_{sym}$) of the symmetry energy for selected values of $x$ can be read from ~\tabref{xvslsymksym}.

\begin{table}
\centering
\begin{tabular}{|c|c|c|}
\hline\hline
x&L(MeV)&K$_{sym}$(MeV)\\ 
\hline\hline
-2& 152 & 418 \\
-1& 106 & 127 \\
0& 61 & -163\\
1& 15 & -454\\
2& -31& -745\\
\hline\hline
\end{tabular}
\caption{Values for $L$ and $K_{sym}$ coefficients appearing in the Taylor expansion of the symmetry energy around saturation density,
$S(\rho)=S_0+L/3\,u+K_{sym}/18\,u^2+\dots$ with $u=\frac{\rho-\rho_0}{\rho_0}$ and $S_0$=30.6 MeV, for given values of the stiffness parameter $x$.}
\tablab{xvslsymksym}
\end{table}

In the previous version of the model, the radius mean square (rms) of initialized nuclei was determined solely from the position of the 
centroids of the wave function of nucleons. This is however inaccurate for the case of Gaussian-type nucleon 
wave functions of finite width, as used in QMD transport models, leading to an effective larger rms. The appropriate expression reads
\begin{eqnarray}
 \langle r^2 \rangle=\frac{1}{N}\sum_{i=1}^{N} (\langle \vec r \rangle - \vec r_i)^2+\frac{3}{2}L_N,
\eqlab{correctedrms}
\end{eqnarray}
where $L_N$ is the square of the nucleon wave function width, the used convention for the parametrization of the nucleon wave function being
the same as in Ref.~\cite{Hartnack:1997ez}. The difference between the previously used and the appropriate value grows 
with increasing wave function width, reaching about 10$\%$ for values customarily used in transport models in connection
with heavy nuclei. While the impact on pion
multiplicities in central collisions is small, leaving the results of Ref.~\cite{Cozma:2014yna} unchanged, the impact on flow
observables in mid-central and, especially, peripheral collisions is non-negligible. The somewhat larger values for the SE slope
parameter at saturation extracted in Ref~\cite{Cozma:2013sja} are corrected downwards by as much as 25 MeV, bringing the extracted
constraints for the SE slope parameter from elliptic flow in Refs.~\cite{Cozma:2013sja, Russotto:2011hq} closer together. The value
for the wave function width in this study is chosen to be $L_N$=4.33 fm$^2$, guided by the ability of reproducing nuclear density profiles,
particularly towards the surface of the nucleus.

In contrast to previous versions of the model, the pion is also associated a finite width wave function, which is introduced
for consistency reasons in order to evaluate the pion-nucleon Coulomb and density dependent strong interactions in the same fashion
as their nucleon-nucleon counterpart. The value of the square of wave function width of the pion is set to half of that of the nucleon,
$L_\pi$=0.5\,$L_N$, 
which is a close approximation of the experimentally measured squared ratio of their charge radii~\cite{Olive:2014aa}. Additionally, the 
strength of the Coulomb interaction has been slightly adjusted  (decreased by 10$\%$ compared to its standard value) in order to reproduce 
more closely, than in previous versions of the model, the Coulomb binding energy contribution to the empirical mass formula, which 
for $^{197}$Au is approximately 3.72 MeV/nucleon~\cite{Liu:2011ama}. This step is justified by the implicit dependence of the 
Coulomb interaction on the value of the wave function width of nucleons (and pions). The impact of this modification on pions is 
non-negligible, as will be shown in \secref{pipotobs}, given its effective isovector nature, leading to lower values for both the 
PMR and the PAPTR. The value of the elliptic flow ratio of neutrons and protons is however only slightly modified.

Most of the results presented in this article have been obtained by enforcing conservation of the total energy of the
system during a heavy-ion collision, by including potential energies in the energy conservation constraint
imposed when determining the final state of a 2-body scattering, decay or absorption process,
\begin{eqnarray}
 \sum_j \sqrt{p_j^2+m_j^2}+U_j&=&\sum_i \sqrt{p_i^2+m_i^2}+U_i,
\end{eqnarray}
both indexes running over all particles present in the system and corresponding, from left to right, to the final and initial states of an 
elementary reaction. This scenario has been referred to as the ``global energy
conservation'' (GEC) scenario in~\cite{Cozma:2014yna}. Additionally, the ``local energy conservation'' (LEC) and 
``vacuum energy conservation'' (VEC) scenarios have been introduced. They correspond to the situation when only the potential
energies of the particles directly involved in the 2-body scattering, decay or absorption process are accounted for
in the energy conservation constraint and when the potential energies of particles in the medium are ignored 
in the collision term, respectively. For further details about these approximations the reader is referred to Ref.~\cite{Cozma:2014yna}.

It will prove useful to mention the used Ansatz for the potential of $\Delta$(1232) and heavier baryonic resonances,
derived under the assumption that it is given by the weighted average of that of neutrons and protons, the weight for each
charge state being equal to the square of the Clebsch-Gordon coefficient for isospin coupling in the process 
$\Delta\rightarrow \pi N$~\cite{Li:2002yda}. It can be cast in the following form, 
\begin{eqnarray}
 \begin{array}{lcr}
V_{\Delta^-}&=& V_N+(3/2)\,V_v\phantom{,}\\
V_{\Delta^0}&=&V_N+(1/2)\,V_v\phantom{,}\\
V_{\Delta^+}&=&V_N-(1/2)\,V_v\phantom{,}\\
V_{\Delta^{++}}&=&V_N-(3/2)\,V_v,
\eqlab{choicedeltapot2}
\end{array}
\end{eqnarray}
where $V_N$ and $V_v$  are the isoscalar nucleon potential and the difference between the potentials of two neighboring isospin
partners respectively. With the assumptions presented above, it can be shown that $V_v$ =$\delta$, with the definition $\delta$=(1/3)($V_n$-$V_p$). 
By varying the magnitude of $V_v$ different scenarios for the strength of the isovector baryon potential
can be explored. The choices $V_v$=-2$\delta$, -$\delta$, 0, $\delta$, 2$\delta$ and 3$\delta$ will be used in this study.
The last choice leads, in the case of a momentum independent potential, to no threshold effects.
The results of this case for the PMR resemble that of transport models that do not take into account the
potential energies in the energy conservation constraint in collision, decay or absorption processes~\cite{Cozma:2014yna}.

\subsection{The Pion Optical Potential}
\seclab{pionoptpot}

\begin{table*}[ht]
 \centering
\begin{tabular}{|l||c|c|c|c||c|c|c|c|c|}
\hline\hline
& $\bar{b}_0$ [$m_\pi^{-1}$] & $\bar{b}_1$ [$m_\pi^{-1}$] & Re $B_0$[$m_\pi^{-4}$] & Im $B_0$[$m_\pi^{-4}$]
&$\phantom{a}\lambda$\phantom{a}& $c_0$ [$m_\pi^{-3}$] & $c_1$ [$m_\pi^{-3}$] & Re $C_0$[$m_\pi^{-6}$] & Im $C_0$[$m_\pi^{-6}$] \\
\hline\hline
SM-1 & -0.0283 & -0.120 & \phantom{-}0.0\phantom{00} & 0.042\phantom{0}& 1 & 0.223 & 0.250 & \phantom{-}0.0\phantom{00} & 0.10\phantom{0}\\
SM-2 & \phantom{-}0.030\phantom{0} & -0.143 & -0.150 & 0.046\phantom{0} & 1 & 0.210 & 0.180 & \phantom{-}0.11\phantom{0} & 0.09\phantom{0} \\
Batty-1 & -0.017\phantom{0} & -0.130 & -0.048& 0.0475 & 1 & 0.255 & 0.170 & \phantom{-}0.0\phantom{00} & 0.09\phantom{0} \\
Batty-2 & -0.023\phantom{0} & -0.085 & -0.021 & 0.049\phantom{0} & 1 & 0.210 & 0.089 & \phantom{-}0.118 & 0.058 \\
Konijn-2 & \phantom{-}0.025\phantom{0} & -0.094 & -0.265 & 0.0546 & 1 & 0.273 & 0.184 & -0.140 & 0.105 \\
\hline\hline
\end{tabular}
\caption{A small sample of the pion optical potential parameter sets extracted from experimental pionic atom
data available in the literature. The entries in this table are a subset of the ones presented in Table II of Ref.~\cite{Itahashi:1999qb},
the labeling being identical. The selection was made such as to cover as much as possible, with a limited number of parameter sets,
of the range of the $S$- and $P$-wave isoscalar and isovector potential strengths extracted from data. The original
references for these parameter sets are: Ref.~\cite{Seki:1983sh} for SM-1 and SM-2, Ref.~\cite{Batty:1978aa} for Batty-1, 
Ref.~\cite{Batty:1983uv} for Batty-2 and Ref.~\cite{Konijn:1990uy} for Konijn-2.}
\tablab{pionpotparameters}
\end{table*}

Theoretical and experimental studies of the pion-nucleus interactions date back to 1950s. Theoretically
motivated parametrizations of the so called pion optical potentials introduced back then~\cite{Kisslinger:1955zz,Ericson:1966fm}
are still in current use when comparing different versions of the potential derived either theoretically from microscopical
models~\cite{Kaiser:2001bx,Nieves:1993ev,Nieves:1991ye, Doring:2007qi} or extracted from a comparison of effective models 
to experimentally measured data for pion-nucleus scattering~\cite{Seki:1983sh,Seki:1983si,Masutani:1985rq} or properties of pionic
atoms~\cite{Itahashi:1999qb,Geissel:2002ur,Suzuki:2002ae}.  The last mentioned studies have also been motivated by the opportunity
to investigate a possible partial restoration of chiral symmetry in nuclei via a modification of the isovector $S$-wave
$\pi N$ scattering amplitude~\cite{Suzuki:2002ae,Yamazaki:2012zza,Friedman:2007zza,Kienle:2004hq,Kolomeitsev:2002gc}.

A commonly used parametrization for the pion optical potential in the context of studying pionic atoms, 
introduced by Ericson and Ericson~\cite{Ericson:1966fm}, reads
\begin{eqnarray}
\eqlab{pionoptpot1}
V_{opt}(r)&=&\frac{2\pi}{\mu}\Big[\,-q(r)+\,\vec\nabla\,\frac{\alpha(r)}{1+\frac{4}{3}\pi\lambda\alpha(r)}\,\vec\nabla\,\Big]
\end{eqnarray}
where
\begin{eqnarray}
\eqlab{pionoptpot2}
 q(r)&=&\epsilon_1(\bar{b}_0\rho+\bar{b}_1\beta\rho)+\epsilon_2B_0\rho^2, \nonumber\\
\alpha(r)&=&\epsilon_1^{-1}(c_0\rho+c_1\beta\rho)+\epsilon_2^{-1}(C_0\rho^2+C_1\beta\rho^2). \nonumber
\end{eqnarray}
In the above expressions $\mu$ is the reduced mass of the pion-nucleus system and
$\lambda$ is the Lorentz-Lorentz correction parameter which accounts for the impact of short-range 
nucleon-nucleon correlations on the potential. 
The extra parameters are defined as follows: $\epsilon_1$=1+$m_\pi/m_N$, $\epsilon_2$=1+$m_\pi/2m_N$, with
$m_\pi$ and $m_N$ the $\pi$-meson and nucleon masses respectively. 
Coordinate dependence of the potentials enters through the expressions for the density
$\rho$ and the isospin asymmetry $\beta$. The parameters $\bar{b}_0$, $\bar{b}_1$ and $B_0$ determine the strength of
the $S$-wave part of the interaction, while the $P$-wave term is described by the ones labeled $c_0$, $c_1$, $C_0$ and $C_1$.
Parameters denoted by capital letters can have both a real and an imaginary part while the others are real. During the last decades,
many sets for the optical potential parameter values have been extracted by fitting available experimental data,
mainly pionic atom properties and pion-nucleus scattering cross-sections. Some differences
do however exist between the many available sets for each parameter and are understood as being due to differences in the fitting procedure, 
some small correction terms ($\it e.g.$ angular transformation terms, Pauli blocking and Fermi averaging~\cite{Seki:1983si}) 
included or omitted by the various analyses or somewhat different experimental data sets.
A limited number of such parameter sets, which were used in the present study, are presented in \tabref{pionpotparameters}. 
A more comprehensive list can be found in Ref.~\cite{Itahashi:1999qb} from where the entries 
listed in \tabref{pionpotparameters} were selected.

As pointed out by several authors~\cite{Seki:1983sh,Seki:1983si,GarciaRecio:1988fg,Chen:1993px}, the density regimes
probed in pionic atoms and elastic pion-nucleus scattering experiments are 0.5-0.75$\rho_0$ and 0.0-0.5$\rho_0$,
respectively. For the pionic atoms case also the momentum of the pion is drastically limited to p$<$0.050 GeV/c (or equivalently
pion kinetic energies $\omega<$ 9 MeV). Extrapolating the  pion potentials to values of these two variables probed in heavy-ion collisions of impact energies
in the range 200-500 MeV/nucleon (0.0-2.5$\rho_0$ for density and 0.0-0.3 GeV/c for pion momentum) leads to unavoidable inaccuracies
which originate from the noted differences in the potential parameter values and may be viewed as model dependence. 
An attempt will be made to estimate its magnitude by determining the observables of interest for different choices of the pionic potentials.

Alternatively, the pion optical potential can be determined theoretically within the framework of effective hadronic models.
Starting from basic interaction terms for the $\pi$NN, $\pi$N$\Delta$ and in some cases also $\pi$NN$^*$(1440) vertices one can
determine the pion potential by computing the in-medium pion self-energy in a perturbative approach, the lowest order corresponding
to a linear density approximation when the energy dependence of the interaction is neglected. Models that go beyond the lowest
order in density are available in the literature, both for $S$-wave~
\cite{GarciaRecio:1988bs,GarciaRecio:1988fg,Nieves:1993ev,Nieves:1991ye,Doring:2007qi} 
and $P$-wave~\cite{GarciaRecio:1989xa,Nieves:1993ev,Nieves:1991ye} components of the potential. Their validity
is restricted to subsaturation densities and for the most sophisticated ones~\cite{Nieves:1991ye, Doring:2007qi} also
to rather low values of the pion kinetic energy, $\omega<$ 50 MeV. Problems similar to the ones noted above occur
also when attempting to use these theoretically determined potentials in simulations of intermediate-energy heavy-ion collisions.

\begin{table}[t]
 \centering
\begin{tabular}{|l||c|c|}
 \hline\hline
& $b_0$ [$m_\pi^{-1}$] & $b_1$ [$m_\pi^{-1}$]\\
\hline
Exp &-0.0001$^{+0.0009}_{-0.0021}$ & -0.0885$^{+0.0010}_{-0.0021}$\\
ChPT &0.0076$\pm$0.0031 & -0.0861$\pm$0.0009\\
WT & 0.0 & -0.0790 \\
\hline\hline
\end{tabular}
\caption{Free-space values of the isoscalar and isovector strengths of the $\pi N$ 
center-of-mass scattering amplitudes, $b_0$ and $b_1$.}
\tablab{vacuumpin}
\end{table}

Extrapolations of the empirically or theoretically derived potentials to pion kinetic energies and densities higher than the ones they are
appropriate for must proceed with care. The treatment of the $S$- and $P$- wave components is necessarily different. The case of  the
$S$-wave potential will be considered at first. As already mentioned, the original goal of studying pionic atoms was to investigate the possibility of a partial restoration of chiral
symmetry inside dense nuclear matter. To this end, the isoscalar and isovector $\pi N$ center-of-mass scattering amplitudes
at threshold in dense nuclear matter, $\bar{b}_0$ and $\bar{b}_1$ need to be compared to their free-space counterparts, $b_0$ and $b_1$.
For these latter ones the extracted values from pionic hydrogen and deuterium X-ray experiments~\cite{Schroder:2001rc} 
and theoretical predictions from chiral perturbation theory (ChPT)~\cite{Baru:2010xn,Baru:2011bw} agree
reasonably well and are more over  well approximated by the Weinberg-Tomozawa (WT) lowest-order chiral limit, as presented
in~\tabref{vacuumpin}.

Initial extraction of the values of $\bar{b}_0$ and $\bar{b}_1$ (at finite density) from experimental data of pionic atoms revealed an unusually large
repulsion in the isoscalar channel~\cite{Batty:1997zp}, as is also evident from~\tabref{pionpotparameters}, which seemed to contradict the expectation
of how chiral symmetry restoration is realized in nature. It was later recognized that, for the isoscalar channel, double-scattering
contributions play a crucial role due to cancellations in the single scattering amplitude that lead to an almost zero isoscalar term at leading order
in the chiral expansion. Consequently, most of the magnitude of $\bar{b}_0$ at finite density originates from the
isovector term due to identical particle correlations in nuclear matter leading to the relation
\begin{eqnarray}
 \bar{b}_0=b_0-\frac{3}{2\pi}\,(b_0^2+2\,b_1^2)\,\Big(\frac{3\pi^2}{2}\rho\Big)^{1/3},
\eqlab{b0eff}
\end{eqnarray}
which needs to be supplement by additional less well understood corrections, such as coherent neutron and proton scattering
lengths or dispersive effects of nuclear pion absorption, in order for a quantitative agreement to be reached~\cite{Krell:1969xn}.

The expressions for the leading order approximation in a pion mass expansion of $b_0$ and $b_1$ in ChPT using the
Weinberg-Tomozawa interaction term~\cite{Weinberg:1966kf,Tomozawa:1966jm},
\begin{eqnarray}
 b_0=0.0 \quad\quad b_1=-\frac{m_\pi}{8\pi\,(1.0+m_\pi/m_N)\,f_\pi^2}\,,
\end{eqnarray}
suggest that in-medium effects on the $\pi N$ scattering amplitude enter via a modification, with density, of the value
of the pion decay constant $f_\pi$, for which the following relation holds for small densities~\cite{Weise:2000xp,Weise:2001sg}
\begin{eqnarray}
 f_\pi^2(\rho)=f_\pi^2(0)-\frac{\sigma\rho}{m_\pi^2},
\end{eqnarray}
where $\sigma=45\pm8$ MeV is the well known pion-nucleon $\sigma$ term, leading to an effective dependence
of $b_1$ on density
\begin{eqnarray}
b_1(\rho)=\frac{b_1}{1-\frac{\sigma \rho}{m_\pi^2\,f_\pi^2}}\simeq \frac{b_1}{1-2.3\rho}.
\eqlab{b1densdep}
\end{eqnarray}
Using these considerations it can be shown that a satisfactory description of pionic atoms and pion-nucleus elastic scattering
can be achieved with values for $b_0$ and $b_1$ in~\eqref{b0eff} compatible with the vacuum ones listed in~\tabref{vacuumpin},
once the energy dependence of the $\pi N$ amplitudes and realistic neutron and protons density profiles inside the nucleus
are also taken into account~\cite{Kolomeitsev:2002gc,Friedman:2004jh,Friedman:2005pt}. It can be additionally shown that, by
enforcing gauge invariance by minimal substitution in the Klein-Gordon equation used to describe pionic atoms, $S$-wave pion potentials
with Re$B_0$=0.0 (in units of $m_\pi^{-4}$) are compatible with experimental data with the magnitude of the imaginary partly largely
unaffected by any of the details of the scenario employed~\cite{Friedman:2014msa}.

\begin{figure*}[htb]
\begin{center}
\begin{minipage}{0.49\textwidth}
\includegraphics[width=16pc]{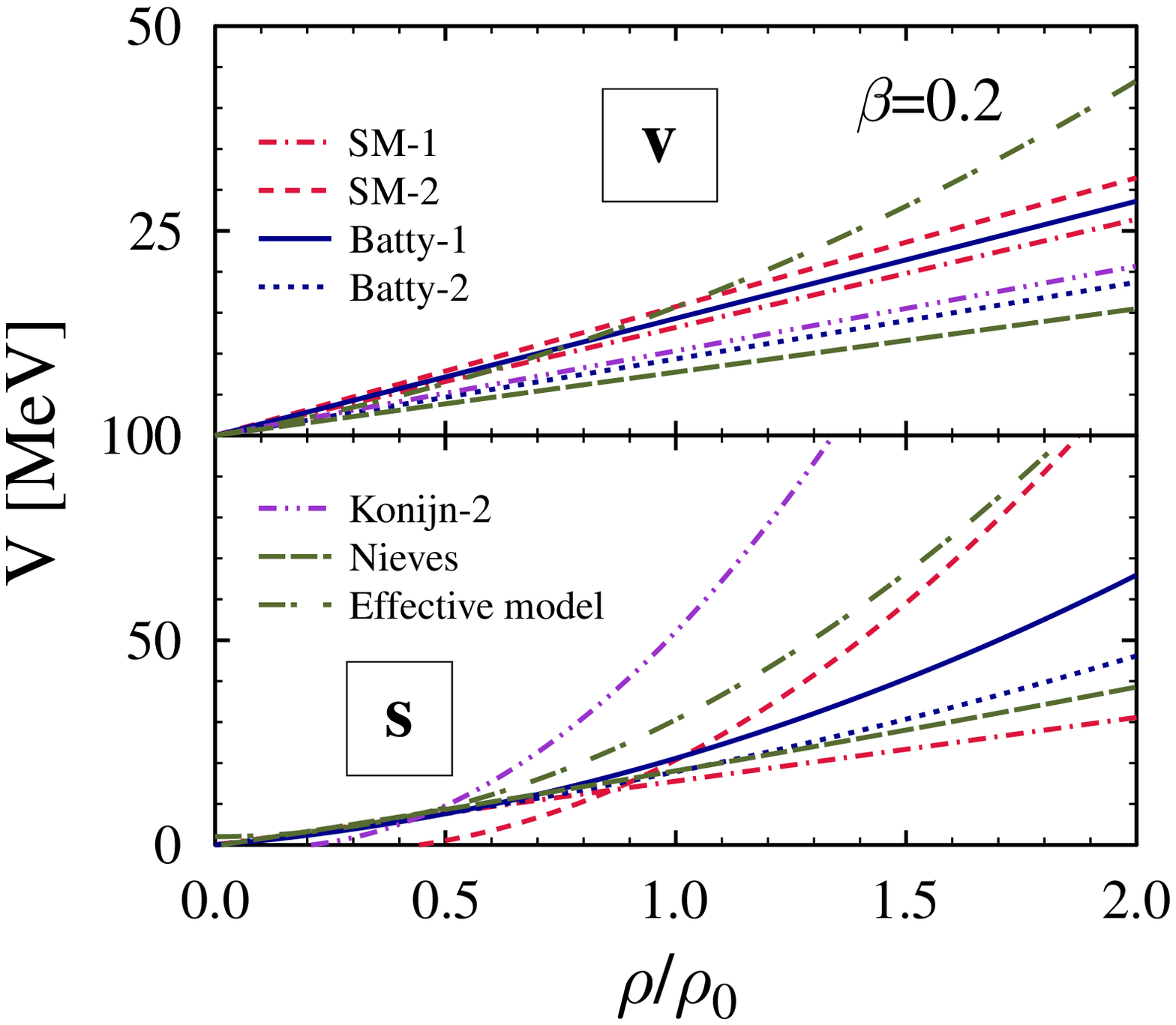}
\end{minipage}
\begin{minipage}{0.49\textwidth}
\includegraphics[width=16pc]{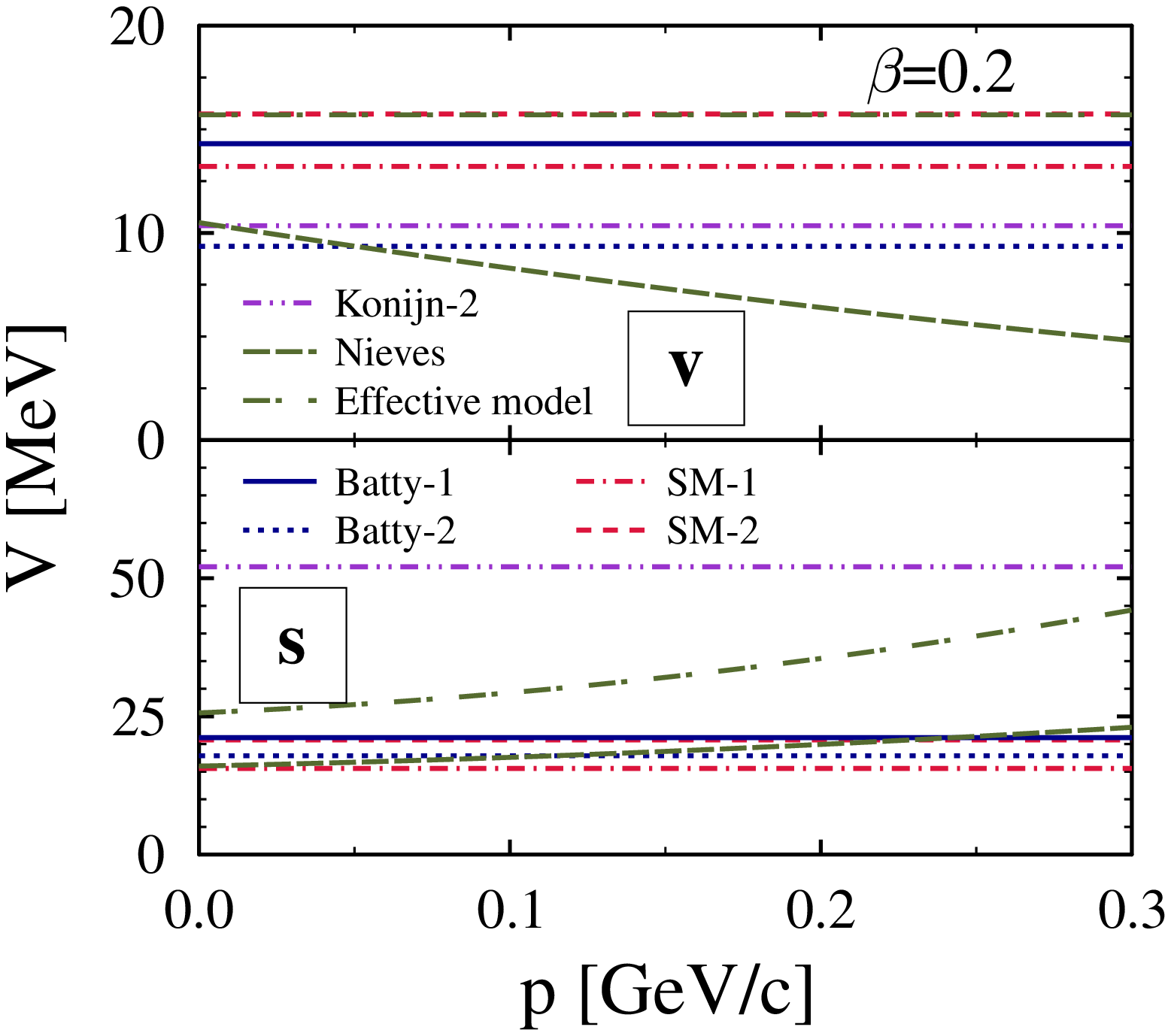}
\end{minipage}
\end{center}
\caption{\figlab{swavepot} (Color online) Density dependence of the pionic $S$-wave potential at fixed momentum, p=0.125 GeV/c (left-hand panels),
and its momentum dependence at fixed density, $\rho/\rho_0$=1.0 (right-hand panels). The total $S$-wave potential has been split into
its isoscalar (bottom panels, labeled ``s'') and isovector (top panels, labeled ``v'') components. In addition to the potentials extracted from pionic atom data, 
the behavior of the theoretical model of Nieves ${\it et~al.}$~\cite{Nieves:1991ye} and of the chiral perturbation theory inspired effective
model discussed in the text are also presented. The value of the isospin asymmetry parameter has been set to $\beta$=0.20, close
to that of the $^{197}$Au nuclei for which the heavy-ion simulations have been performed.}
\end{figure*}

As mentioned in the previous paragraph, the energy dependence of the $S$-wave potential is also important for the description
of experimental data away from threshold (pion-nucleus scattering). In the leading order approximation of ChPT (pion kinetic
energy smaller than pion rest mass) the dominant energy dependence originates from $b_0$. This result is supported by
the energy dependence of empirical free-space $\pi N$ amplitudes that have been extracted from experimental data 
(see for example Ref.~\cite{Arndt:2006bf}), which advocate slopes of the potential parameters $b_0$ and $b_1$ 
of -0.00053 $m_\pi^{-1}/\mathrm{MeV}$ and a negligible one, respectively. Analyses of low-energy pion-nucleus scattering
arrive at a qualitatively identical conclusion~\cite{Seki:1983sh,Seki:1983si,Masutani:1985rq}, additionally presenting evidence
of a dampening of the energy dependence of the isoscalar term of the $S$-wave scattering amplitudes in nuclear matter 
as compared to free space.

Consequently, in the simulations presented in the next section the values
of $b_0$ and its slope were chosen with the conservative requirement of satisfying the experimental 
constraints derived from pion-nucleus scattering for the so called effective isoscalar scattering amplitude 
$\bar{b}_0^{\mathrm{eff}}$~\cite{GarciaRecio:1988fg}: $b_0=-0.010~m_\pi^{-1}$ and $db_0/d E_{kin}=-0.00016~m_\pi^{-1}/\mathrm{MeV}$.
It is defined as: $\bar{b}_0^{\mathrm{eff}}=\bar{b}_0+\rho^{\mathrm{eff}}\,\mathrm{Re}\,B_0$~\cite{Seki:1983sh},
 neglecting small corrections proportional to the ratio between the pion kinetic energy and the mass of the nucleus under consideration, 
with $\rho^{\mathrm{eff}}$ being the effective density at which the potential needs to be evaluated at a given pion kinetic energy.

This approach accounts, even though in a rather qualitative manner, for modifications, induced by the dense medium, 
of the slopes of the energy dependence of the parameters of the potential. The non-zero value of $b_0$, 
different from its free-space value (see~\tabref{vacuumpin}), effectively accounts 
for the omitted corrections in the process of deriving~\eqref{b0eff}~\cite{Krell:1969xn} at finite density. 
The described procedure to account
for the energy dependence of the pion-nucleon scattering amplitudes resembles the approach employed 
in pionic atom studies~\cite{Friedman:2014msa}. The analytical dependence on energy of the theoretical $S$-wave
pion potential of Ref.~\cite{Nieves:1991ye} is however different, leading to a smaller energy slope of $\bar{b}_0$ but
to an energy dependent $\bar{b}_1$ (see the right panel of~\figref{swavepot}).

The following $S$-wave potential will be used in the numerical simulations of heavy-ion collisions, if not otherwise stated. 
For $b_0$ the value of the slope extracted above from experimental pion-nucleus elastic scattering will be used,
while for $b_1$ a linearized approximation of~\eqref{b1densdep}, that is applicable (non-singular) to the entire density
interval probed by intermediate energy heavy-ion collisions, will be employed
\begin{eqnarray}
\eqlab{b0b1final}
b_0(\omega)&=&-0.010 -0.00016\,\omega  \\
b_1(\rho)&=&-0.088\,\Big(1+\frac{0.6116}{b_1}\,\frac{\rho}{\rho_0}\Big). \nonumber
\end{eqnarray}
Both potential parameters in the expressions above are expressed in units of [$m_\pi^{-1}$], while the kinetic energy of the pion, $\omega$,
is expressed in units of [MeV].
The values of the parameters entering in~\eqref{pionoptpot1} are determined from~\eqref{b0eff} for $\bar{b}_0$ and $\bar{b}_1=b_1(\rho)$
for $\bar{b}_1$. It needs to be stressed that the above choice for $b_1$ (together with the one in~\eqref{b1densdep}) may not be a very good
approximation far away from the low density region, however neglecting the density dependence
of $b_1$ completely may arguably be a worse approximation. The precise (realistic) dependence on density of $\bar{b}_1$ for the entire 
density domain of interest in this study is presently unknown and any extrapolation of low density
ChPT results of the type of the one in~\eqref{b0b1final} introduces unavoidable uncertainties in the results. 

\begin{figure*}[htb]
\begin{center}
\begin{minipage}{0.49\textwidth}
\includegraphics[width=16pc]{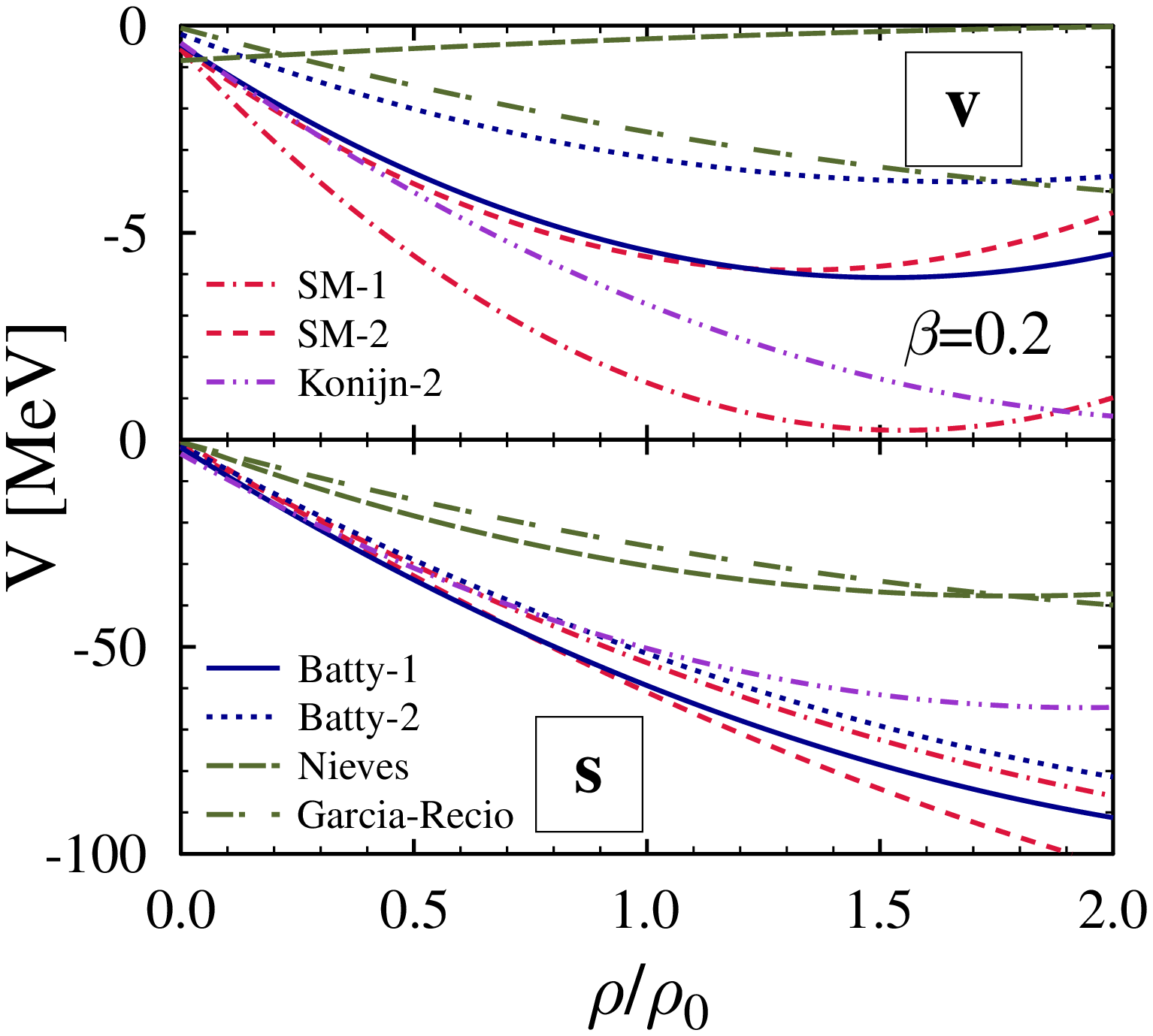}
\end{minipage}
\begin{minipage}{0.49\textwidth}
\includegraphics[width=16pc]{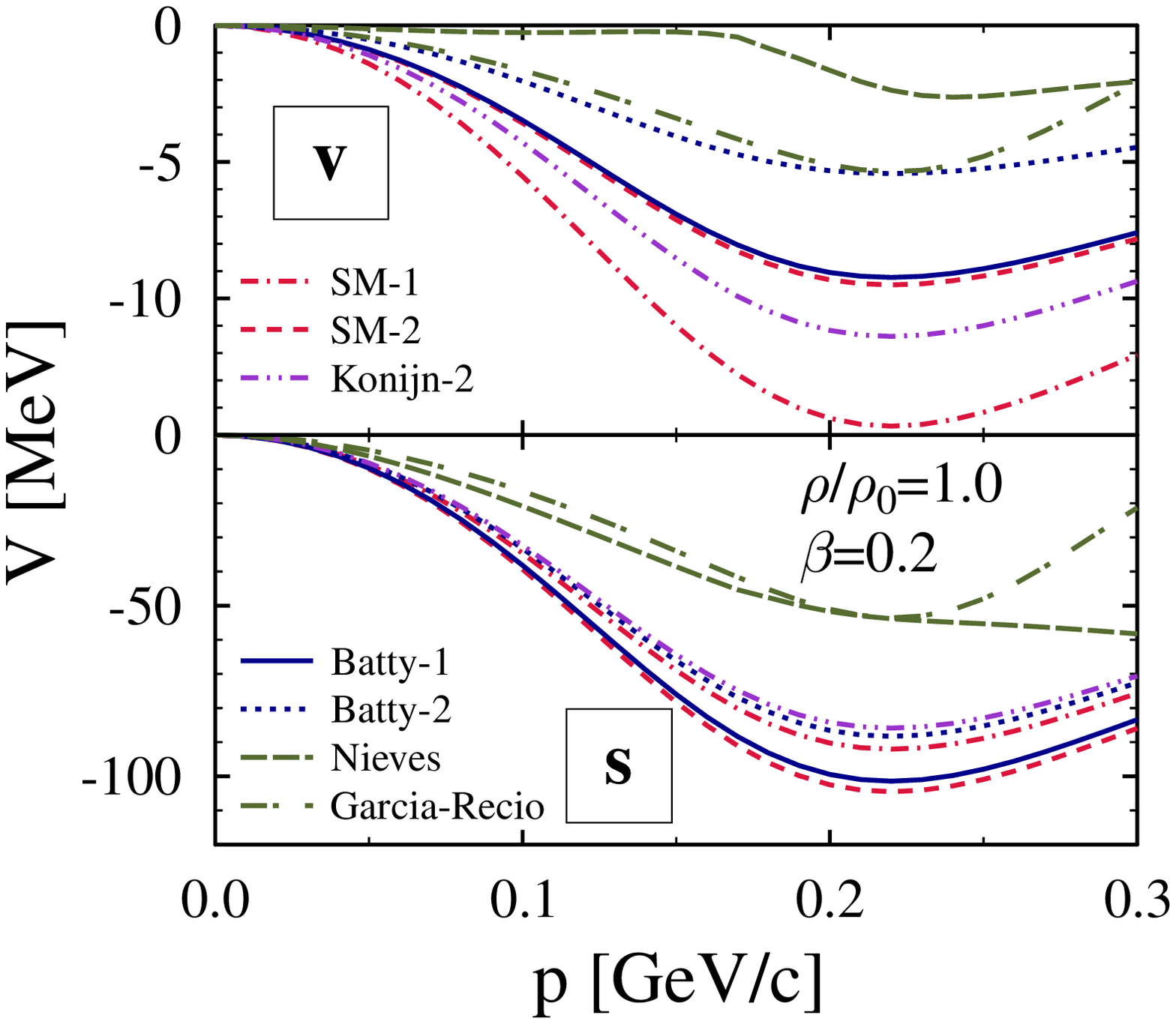}
\end{minipage}
\end{center}
\caption{\figlab{pwavepot}(Color online) The same as in~\figref{swavepot}, but for the $P$-wave potential. In this case the results
of two theoretical models are shown, namely the one of Nieves$~{\it et~al.}$~\cite{Nieves:1991ye} and that of
Garcia-Recio$~{\it et~al.}$~\cite{GarciaRecio:1989xa}. As discussed in the text the momentum dependence of the $P$-wave potentials extracted
from pionic atoms data has been extrapolated using the momentum dependence of the latter theoretical model. The presented results
correspond to the case of \underline{uniform} nuclear matter of given density and isospin asymmetry $\beta$=0.2. Consequently, the
density gradient term that appears in~\eqref{pionoptpot1} does not contribute.}
\end{figure*}

The discussion of the $S$-wave potential is concluded by presenting, in~\figref{swavepot}, the density (left panel) and
momentum (right panel) dependence of the empirical $S$-wave potentials of ~\tabref{pionpotparameters}, of the
theoretical model of Nieves$~{\it et~al.}$~\cite{Nieves:1991ye} and the effective model presented above. 
The isoscalar and to a lesser extent also the isovector components are compatible with each
other in the density region probed in pionic atom experiments; their strengths differ however substantially
in the supra-saturation region. Only the theoretical and effective models for the $S$-wave pion potential present
a momentum dependence, which are rather different from each other, the latter having a stronger dependence
in the isoscalar channel and none for the isovector case.

Turning to the $P$-wave potential, one can notice from~\tabref{pionpotparameters} that, by using the concept of 
effective density~\cite{Seki:1983sh}, the various sets of potentials extracted from pionic atoms present isoscalar components
of similar strength, while for the isovector term $c_1$ the strength varies, in absolute magnitude, by a factor of three and
consequently also the ratio of the strengths of the isovector and isoscalar components varies within a similar range. This
is visible in the left panel of~\figref{pwavepot} where the density dependence of the isoscalar and isovector $P$-wave potentials
at a value of the pion momentum p=0.125 GeV is presented. Additionally, two theoretical pion $P$-wave pion potentials are also
depicted, whose strength is systematically smaller than that of the empirical ones extracted from pionic atoms data.
It is expected that the mentioned differences may have an important impact on observables that probe the isovector
part of the interaction. The strong dependence of the $P$-wave potential on pion momentum and the rather important dependence
of the average pion momentum on the isospin in heavy-ion collisions lead to an effective isovector behavior also 
of the isoscalar $P$-wave terms. This fact stresses the importance of an accurate knowledge of the pion potential.

A realistic dependence on momentum of the $P$-wave potential is therefore crucial, as will be shown for the observables
of interest in~\secref{pipotobs}. The evident $p^2$ dependence 
from~\eqref{pionoptpot1} is only valid at small pion kinetic energies and far away from the position of the pole masses
of excitable baryonic resonances. This requirement is fulfilled in the case of pionic atoms. As the energy is increased and the 
lowest lying resonance, $\Delta$(1232), is excited, the dependence of the pion $P$-wave potential on momentum is modified, 
influenced primarily by the energy dependence of the decay width of the resonance in question.
A precise energy dependence of the potential can currently only be inferred from models that can determine the pion self-energy
in nuclear matter for a wide enough kinetic energy range. In this respect, the theoretical model of Ref.~\cite{GarciaRecio:1989xa}, based
on a local approximation of the delta-hole model, has allowed a good description of pion nucleus scattering up to kinetic energies
of the incident pion of about 300 MeV. While more sophisticated models do exist in the literature~\cite{Nieves:1991ye}, they have
a limited range of applicability (pion kinetic energy $\omega \leq$50 MeV) and consequently present
an (unrealistic) increase in strength with $\omega$ even for invariant mass values above the position of the $\Delta$(1232) resonance. 
Extrapolations of such a potential, above its range of applicability, by using three-level type models for
the pion-self energy in nuclear matter, as the one proposed in Ref.~\cite{Buss:2006vh}, and used recently
in~\cite{Guo:2014fba}, are considered here also inaccurate since the width of the $\Delta$(1232), not just its energy dependence, is
completely neglected in these cases. The energy dependence of the $P$-wave potential
derived in Ref.~\cite{GarciaRecio:1989xa} is adopted in this study. In practice, this is achieved by multiplying
 the $P$-wave part of the potential of~\eqref{pionoptpot1} by the form factor,
\begin{eqnarray}
 f(p^2)=\frac{1.0-p_{eff}^2/\Lambda_1^2+p_{eff}^4/\Lambda_2^4}{1.0-p^2/\Lambda_1^2+p^4/\Lambda_2^4},
\end{eqnarray}
with $\Lambda_1=0.55~\mathrm{GeV}$ and $\Lambda_2=0.22~\mathrm{GeV}$. The expression in the numerator ensures that for a value of
the pion momentum equal to that of the average one in a 2p state of a heavy pionic atom ($p_{\mathrm{eff}}=0.05$ GeV) 
the strength of the potential as extracted from pionic atom measurements is reproduced. In~\secref{pipotobs}, in order to test the
sensitivity of the observables to various isoscalar and isovector strengths of the $P$-wave potential, results will be presented for all
potentials listed in~\tabref{pionpotparameters} and also for the potential of Nieves$~{\it et~al.}$~\cite{Nieves:1991ye}.

The momentum dependence of the pion P wave potentials of~\tabref{pionpotparameters} and of the
theoretical models of Nieves$~{\it et~al.}$~\cite{Nieves:1991ye} and Garcia-Recio~\cite{GarciaRecio:1989xa}
is depicted in the right panel of~\figref{pwavepot} for the isoscalar (s) and isovector (v) components
separately. The features described in the previous paragraph are readily observable and, as in the case of
the density dependence, the theoretical models exhibit weaker attraction than the empirical ones.

Due to the dependence of the density and isospin asymmetry parameters on the spatial coordinates, the gradient operator in
~\eqref{pionoptpot1} leads, besides the term proportional to $p^2$, also to terms in the potential proportional to 
$\vec{p}\cdot\vec\nabla \rho$ and $\vec{p}\cdot\vec\nabla \beta$. They can in principle be of relevance in the study
of  pionic atoms~\cite{Ericson:1966fm} since in this case the pion probes mostly the region close to the surface of the nucleus. 
Theoretical investigations on this topic make use of density profiles of nuclei that
lead to constant isospin asymmetry~\cite{Itahashi:1999qb,Yamazaki:2012zza} within the nucleus and 
consequently the isospin asymmetry gradient term does not contribute. In order to be consistent with the studies that have
lead to the pion potentials of~\tabref{pionpotparameters} terms in the potential proportional to
$\vec{p}\cdot\vec\nabla \beta$ will be neglected in the following, keeping however those proportional to the density gradient. 

\begin{figure}
 \includegraphics[width=16pc]{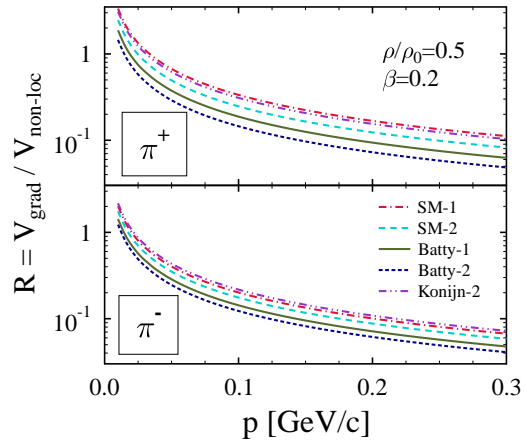}
\caption{\figlab{gradpwavepot} Ratio $R$ of the density gradient component of the $P$-wave pion potential (V$_{grad}$)
versus the component proportional to $p^2$ (V$_{non-loc}$) for the $\pi^-$ (bottom) and $\pi^+$ (top) mesons.  
Results for various choices of the $P$-wave potential parameter set (see~\tabref{pionpotparameters}) are presented.
The calculation has been performed for a $^{197}$Au nucleus whose density profile can be parametrized 
by the simple expression $\rho(r)=\rho_0/(1+exp[(r-R)/a])$, with $\rho_0$=0.165 fm$^{-3}$, R=6.40 fm and a=0.60 fm. 
The value of the coordinate $r$ is chosen such as to maximize the magnitude of the density gradient,
which for the chosen parametrization occurs at the location at which $\rho$=$\rho_0$/2 and hence $r$=R. The spread
of the results for the chosen parameters sets of the P wave potential is essentially given by the variation of the magnitude
of Im$C_0$ between the different potentials.}
\end{figure}

Their relevance can be inferred from~\figref{gradpwavepot}, in which the ratio of the strengths of density gradient terms
in the potential and of the $p^2$ term of  $P$-wave potential as a function of momentum of the pion, for several choices
of the $P$-wave potential parameter set and a modulus of the radius vector for which the density gradient is maximum, 
are presented. The calculations have been performed for the case of a $^{197}$Au
nucleus with the density profile as specified in the caption of~\figref{gradpwavepot} and outward radial pion momentum 
orientation. Results for the $\pi^-$ and $\pi^+$ mesons are presented in the bottom and top panels respectively. It is readily observed that in both cases the relative strength of the gradient term potential is stronger
at lower momenta, where it becomes the dominant contribution to the total pion $P$-wave potential. At higher momenta this 
relative contribution decreases to about or even below 10$\%$. Additionally, a variation of the relative strength
of the density gradient term within a factor of 2 between the different choices for the $P$-wave potential is observed, which
can be predominantly linked to the value of the Im$C_0$ parameter (see~\tabref{pionpotparameters}), which sets the strength
of the two-body pion absorption processes. In the case of heavy-ion collisions, density gradients of comparable magnitude with
the ones encountered in the skin of nuclei are produced over a wider range of density values. It is thus mandatory to investigate the impact
of the density gradient term of the $P$-wave potential on pionic observables.

\begin{figure}
 \includegraphics[width=16pc]{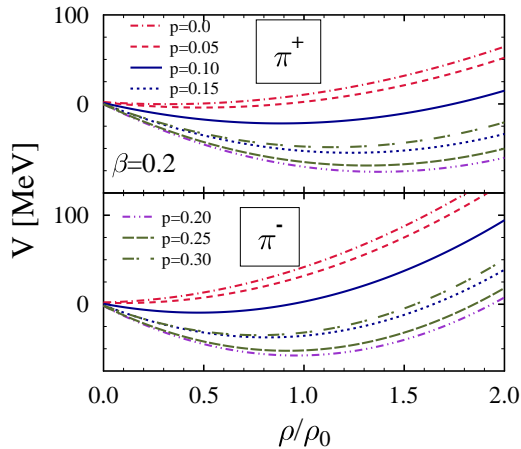}
\caption{\figlab{spwavepot} (Color online) Total $\pi^{-}$(lower panel) and $\pi^{+}$ (upper panel) potentials in 
\underline{uniform} nuclear matter of isospin asymmetry $\beta$=0.2 for several values of the pion momentum (expressed in GeV/c).
The chiral perturbation theory inspired effective model was chosen for the $S$-wave part, while for the $P$-wave part the potential labeled
Batty-1 has been selected. The other possible combinations yield qualitatively the same behavior. The repulsion generated by 
the density dependence of the $S$-wave isovector strength $\overline{b}_1$ overcomes eventually the attraction in the $P$-wave 
channel leading to a transition from a net attractive to a net repulsive potential for ever increasing, 
with the pion momentum, values of the density.
}
\end{figure}

This section is concluded by presenting, in~\figref{spwavepot}, the $\pi^-$ (bottom panel) and $\pi^+$ (top panel) 
total S+P pion potential in uniform nuclear matter as a function of density for various values of the pion momentum $p$.
The effective model is chosen for the $S$-wave part and the Batty-1 parameter set of~\tabref{pionpotparameters} for
the $P$-wave component. The total potential is repulsive for small values of the pion momentum irrespective of density. 
At higher momenta, the potential becomes attractive; however, as the density increases, the repulsive
$S$-wave part prevails resulting again in a net repulsive interaction. Due to the isovector component, the potentials
of the $\pi^-$ and $\pi^+$, while showing qualitative similarities, differ in strength by non-negligible amounts with
foreseeable impact on isovector observables.

\begin{figure}
 \includegraphics[width=16.0pc]{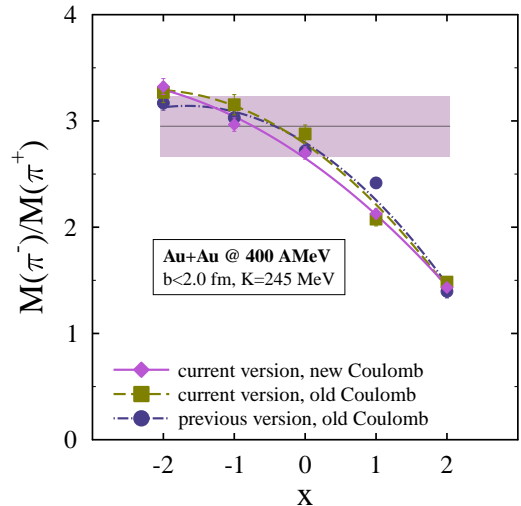}
\caption{\figlab{oldvsnewmodel}(Color online) Impact of the modifications of the model described in~\secref{transpmodel} on
the multiplicity ratio of $\pi^-$ and $\pi^+$ mesons as compared with the result of Ref.~\cite{Cozma:2014yna}. The ordinate
$x$ parametrizes the stiffness of the symmetry energy, negative and positive values corresponding to a stiff and a soft
asy-EoS, respectively. The horizontal band depicts the experimental result of the FOPI Collaboration~\cite{Reisdorf:2010aa}.}
\end{figure}

\section{Pion potential and pion observables}
\seclab{pipotobs}

\begin{figure*}[htb]
\begin{center}
\begin{minipage}{0.49\textwidth}
\includegraphics[width=16.0pc]{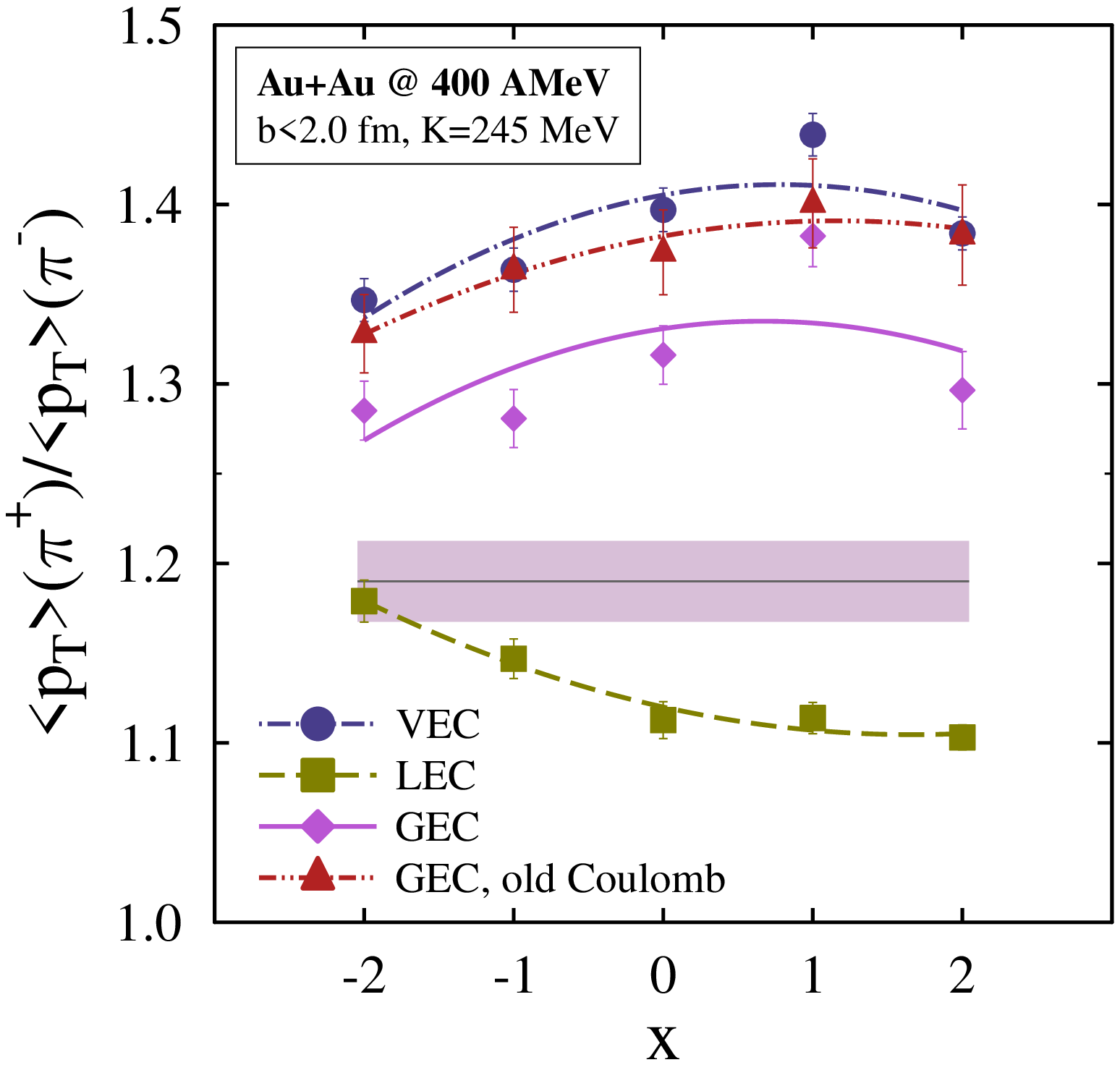}
\end{minipage}
\begin{minipage}{0.49\textwidth}
\includegraphics[width=16.0pc]{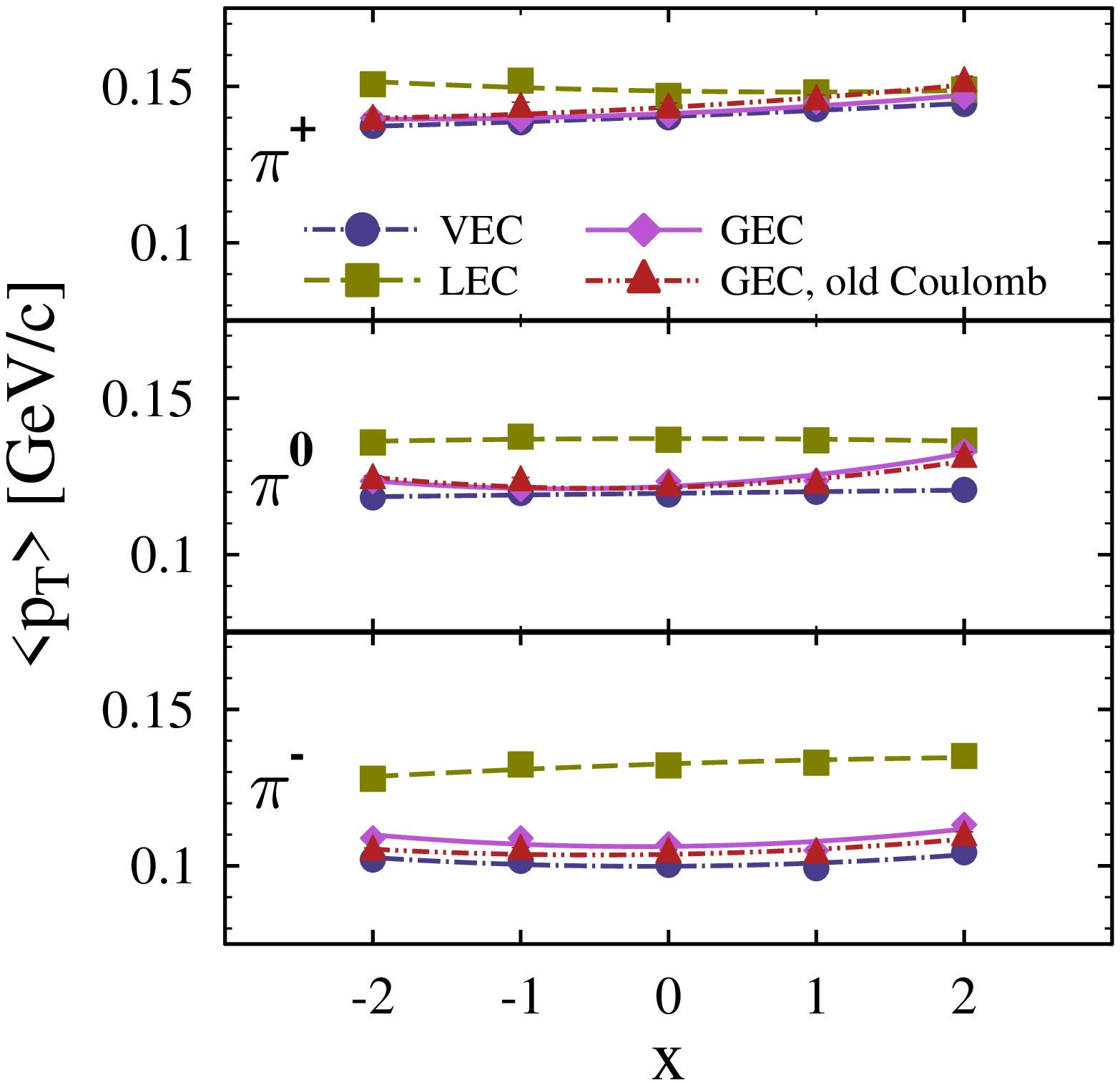}
\end{minipage}
\end{center}
\caption{\figlab{avptvlgecs} (Color online) Average $p_T$ ratio of charged pions (left panel) and average $p_T$ of all charge
states of the $\pi$ meson (right panel) as a function of the stiffness parameter $x$ in central $^{197}$Au+$^{197}$Au
collision at an impact energy of 400 MeV/nucleon. Results for the VEC (dashed-dotted curves), 
LEC (dashed curves) and GEC scenarios are presented. For the case of the GEC scenario results for two strengths of the Coulomb
interaction are shown: the one of Ref.~\cite{Cozma:2014yna} (dashed-double-dotted curves) and the one used in this study (full curves).
The FOPI experimental result for the PAPTR~\cite{Reisdorf:2006ie} is depicted by a horizontal band.  }
\end{figure*}

The impact that the small modifications to the transport model, as described in~\secref{transpmodel},
have on the PMR is presented in~\figref{oldvsnewmodel}. A comparison between the PMR
in central $^{197}$Au+$^{197}$Au central collisions at an impact energy of 400 MeV/nucleon for the case of the 
previous version of the model~\cite{Cozma:2014yna} (dashed-dotted curve) and the current one when first 
the corrected expression of~\eqref{correctedrms} for rms is used (dashed curve), and then also the strength of the Coulomb 
is adjusted to match its contribution to the binding energy as provided by the empirical nuclear mass formula (full curve), 
as described in~\secref{transpmodel}, is presented. For each of the calculations presented the GEC scenario has been adopted,
the isovector $\Delta$(1232) potential strength has been set to $V_v$=$\delta$ and the pion potential has been switched off.
The rest of the model parameters are set to values as described in \secref{transpmodel}. The extracted value of the SE stiffness is only marginally impacted
by the correction to the rms formula. On the other hand, the modification of the strength of the Coulomb interaction leads
to a stiffer asy-EoS, the increase of the extracted slope $L$ of SE being of the order of 15 MeV (see~\tabref{xvslsymksym} 
for the connection between $x$ and $L$).This is comparable with the precision with which
this parameter is extracted from the most recent elliptical flow results of the ASYEOS Collaboration~\cite{Russotto:2016ucm}.

The study reported in Ref.~\cite{Cozma:2014yna} has demonstrated that the ratio of charged pions multiplicity 
is equally sensitive to both the stiffness of the SE and the strength of the isovector $\Delta$(1232) potential in nuclear matter. 
Consequently, constraints for the slope $L$ of the SE at saturation cannot be extracted unambiguously 
without a proper knowledge of the latter. The only solution out of this problem, given that no information about
the isovector $\Delta$(1232) potential is available from either theory or experiment, is to enlarge the set of observables
from which the unknown parameters of the model are extracted. Obvious candidates are the average final momenta
(or kinetic energies) of charged pions. To isolate the isovector signal, similarly to the case of multiplicities, it will prove useful to construct 
their ratio. In addition to multiplicities, the FOPI experiment has also measured the final transverse momenta
of pions and results for the ratio of average $p_T$ of $\pi^+$ and $\pi^-$ are available in the literature for several
systems and impact kinetic energy equal to or higher than 400 MeV/nucleon~\cite{Reisdorf:2006ie}. Consequently,
results and comparisons with available experimental data~\cite{Reisdorf:2006ie,Reisdorf:2015aa}, 
for average transverse momenta of pions and their ratio, will also be presented, where considered useful.

The impact that the various energy conservation scenarios, introduced in Ref.~\cite{Cozma:2014yna} and briefly described in
~\secref{transpmodel}, has on average $p_T$ of pions and their ratio is presented in~\figref{avptvlgecs}. Its left-hand side 
panel presents the impact of the VEC, LEC and GEC scenarios on the pion average $p_T$ ratio (PAPTR). VEC and GEC scenario
simulations reveal values of PAPTR that overshoot the experimental FOPI result~\cite{Reisdorf:2006ie} by 10-20$\%$. On the
other hand the LEC scenario leads to PAPTR values below their experimental counterpart by at most 10$\%$. The impact of local
energy conservation (as compared to VEC) is therefore much more pronounced for PAPTR than for multiplicity ratios, 
while the difference between LEC and GEC scenarios is equally dramatic for these two observables. Decreasing the strength of
the Coulomb interaction by 10$\%$(previous vs. current versions of the model) results in a reduction, in agreement with expectations, 
of the PAPTR by about 5$\%$. A moderate dependence of PAPTR on the SE stiffness is also demonstrated, a softer 
asy-EoS leading to a higher PAPTR for the VEC and GEC scenarios and the opposite for LEC.

\begin{figure*}[htb]
\begin{center}
\begin{minipage}{0.49\textwidth}
\includegraphics[width=16.0pc]{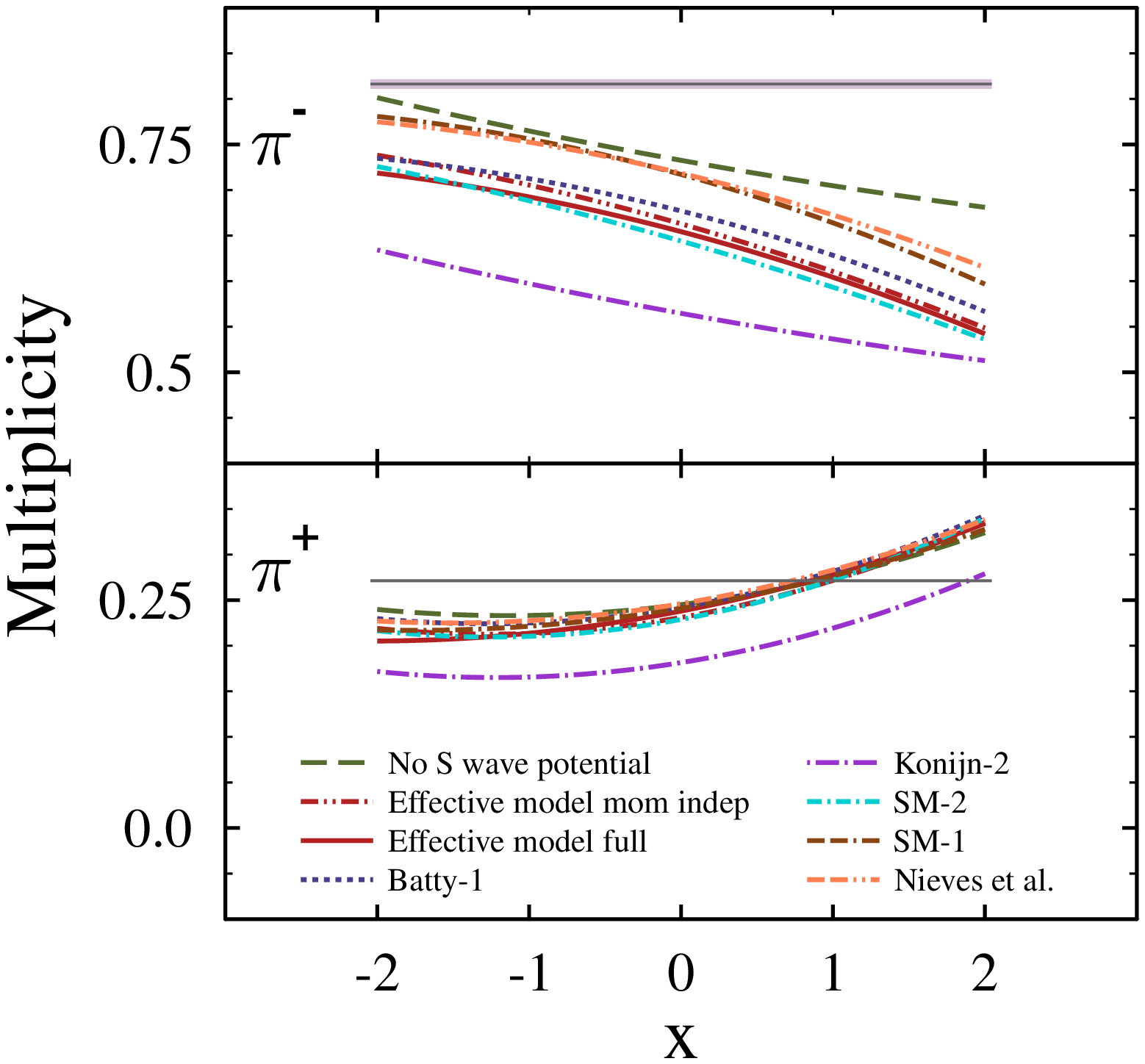}
\end{minipage}
\begin{minipage}{0.49\textwidth}
\includegraphics[width=16.0pc]{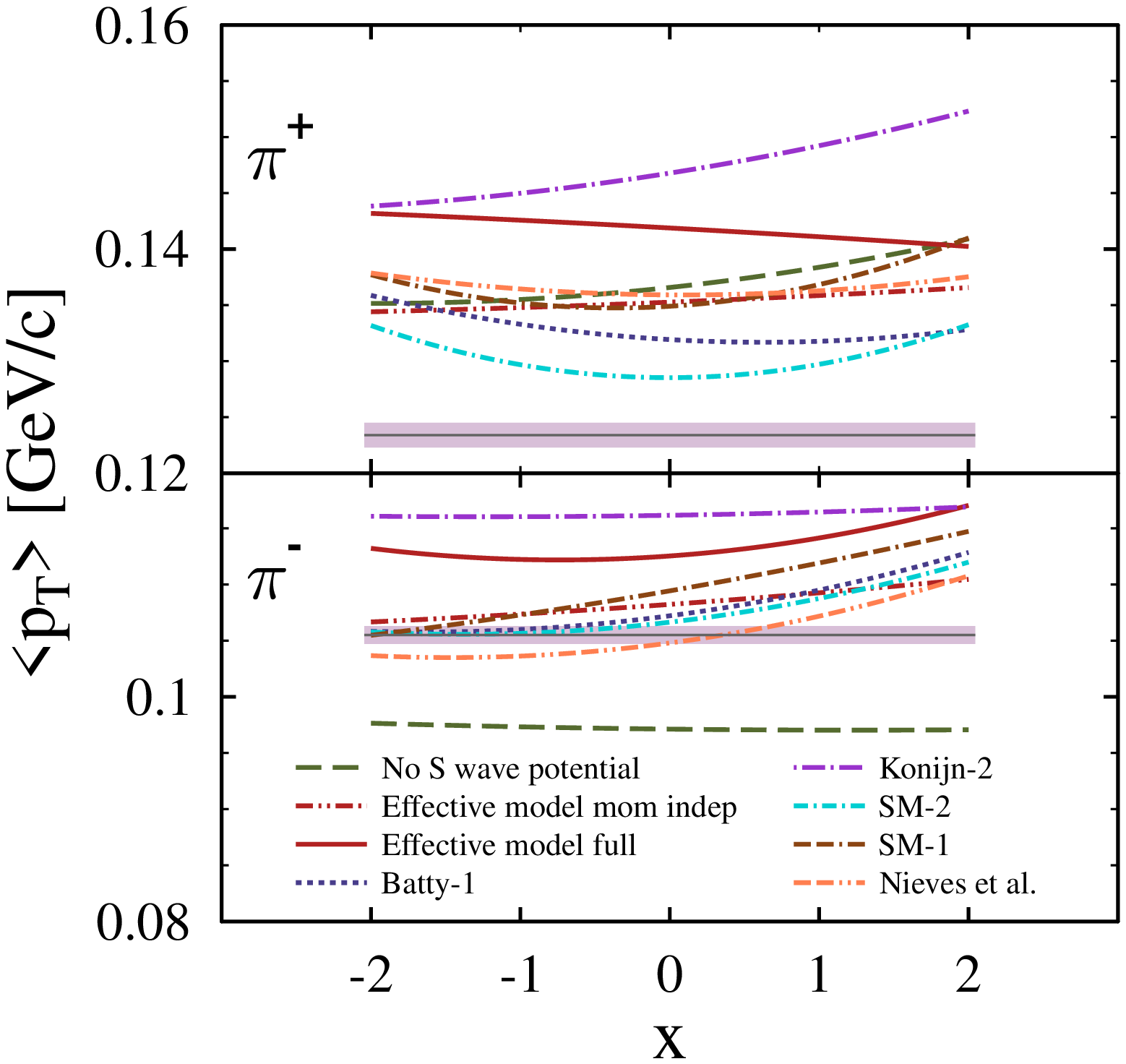}
\end{minipage}
\end{center}
\caption{\figlab{swavepotimpact} (Color online)  Impact of the $S$-wave pion potential on pion multiplicities (left panel)
and pion average transverse momenta (right panel) in mid-central collisions (3.35 fm $<$ b $<$ 6.00 fm) of
$^{197}$Au+$^{197}$Au at an incident energy of 400 MeV/nucleon for various choices of the $S$-wave potential, as presented
in \secref{pionoptpot}. The following kinematical cuts have been applied: $p_{T}<$ 0.33 GeV/c and $|y|<$1.75. 
The curves labeled ``No S wave potential''
were obtained by omitting any contributions due to the pion optical potential. Experimental data~\cite{Reisdorf:2015aa} are
represented by horizontal bands, with their widths representing only the statistical uncertainties (systematic uncertainties
were not available).}
\end{figure*}

\begin{figure*}[htb]
\begin{center}
\begin{minipage}{0.49\textwidth}
\includegraphics[width=16.0pc]{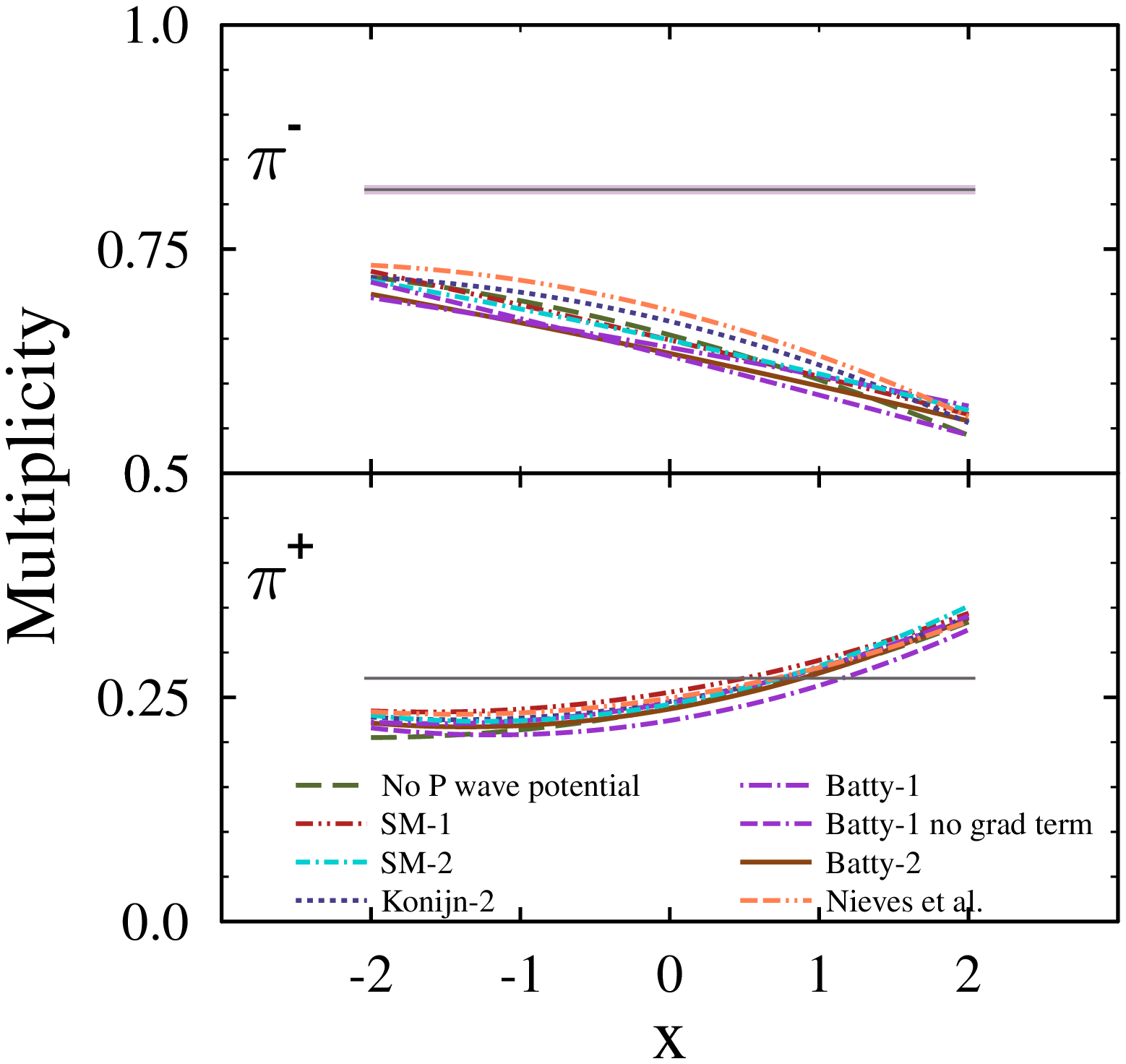}
\end{minipage}
\begin{minipage}{0.49\textwidth}
\includegraphics[width=16.0pc]{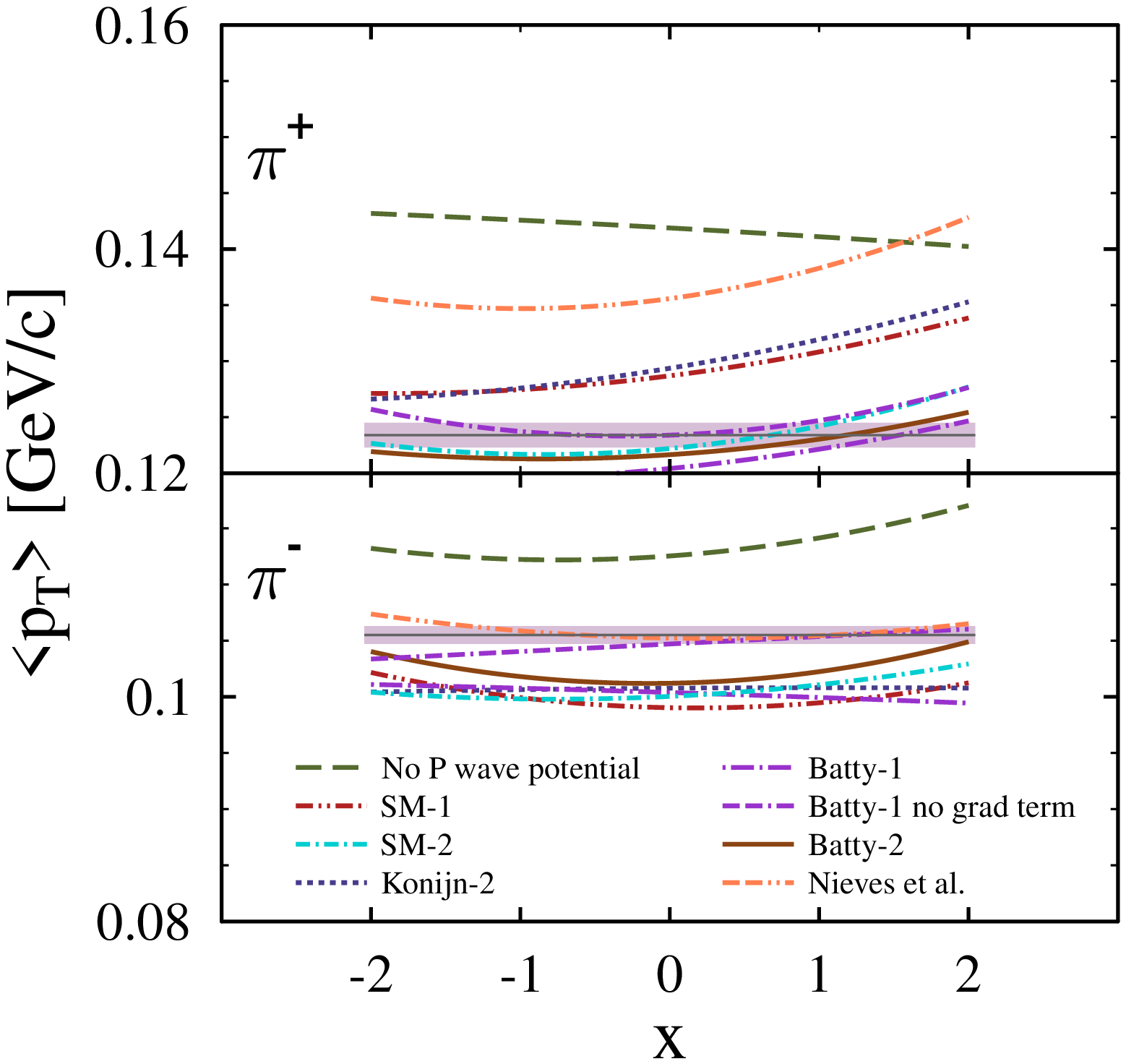}
\end{minipage}
\end{center}
\caption{\figlab{pwavepotimpact} (Color online)  Impact of the $P$-wave pion potential on pion multiplicities (left panel)
and pion average transverse momenta (right panel) in mid-central collisions (3.35 fm $<$ b $<$ 6.00 fm) of
$^{197}$Au+$^{197}$Au at an incident energy of 400 MeV/nucleon for various choices of the $P$-wave potential,
as presented in \secref{pionoptpot}. The following kinematical cuts have been applied: $p_{T}<$ 0.33 GeV/c and $|y|<$1.75. 
The calculations labeled ``No P wave potential''
include the impact of the effective model $S$-wave pion potential alone. The same remark for the experimental values
(horizontal bands), as the one made in the caption of~\figref{swavepotimpact}, holds true.}
\end{figure*}

The right-hand panel of~\figref{avptvlgecs} gives the sensitivity of the average transverse momenta of each of the three
charge pion states to the selected energy conservation scenario. It is shown that the low value of PAPTR in the case of the LEC
scenario originates predominantly from the impact the local conservation of energy (LEC) has on the average 
transverse momentum of the $\pi^-$ meson.
The same holds true for the origin of the differences between the VEC and GEC scenarios and the modification on PAPTR induced
by changing the strength of the Coulomb interaction, even though in this case the changes are much smaller in magnitude.
These observations are on par with the impact of the energy conservation scenarios on pion multiplicities~\cite{Cozma:2014yna}.

The results of the attempt to explain the remaining difference between the GEC result for PAPTR and its experimental value by including
the effect of the pion-nucleus potential will be presented in the following. Since experimental values for pion average transverse momenta were available only
for mid-central collisions (3.35 fm $<$ b $<$ 6.0 fm)~\cite{Reisdorf:2015aa} the study of the impact of pion potentials on pion
observables has been performed for this impact parameter range. To allow a comparison with the experimental data, the following kinematical filter
has been applied to theoretical data: $p_{T}<$ 0.33 GeV/c and $|y|<$1.75. Constraints on the symmetry energy stiffness will however be extracted from
published central collision data (b $<$ 2.0 fm), since only for this case systematical uncertainties have been included
in the estimation of total uncertainties of experimental data.

\begin{figure*}[htb]
\begin{center}
\begin{minipage}{0.49\textwidth}
\includegraphics[width=16.0pc]{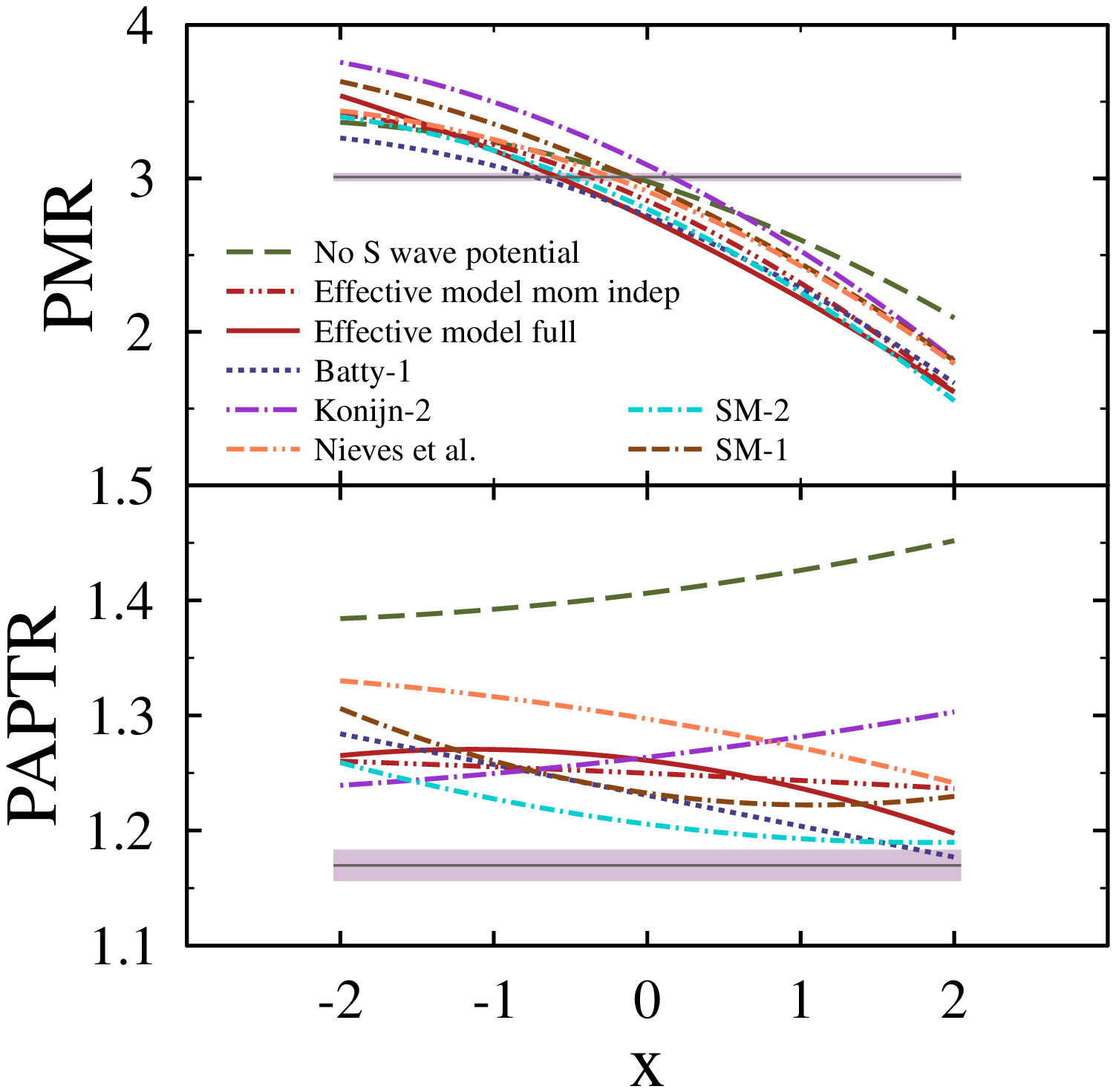}
\end{minipage}
\begin{minipage}{0.49\textwidth}
\includegraphics[width=16.0pc]{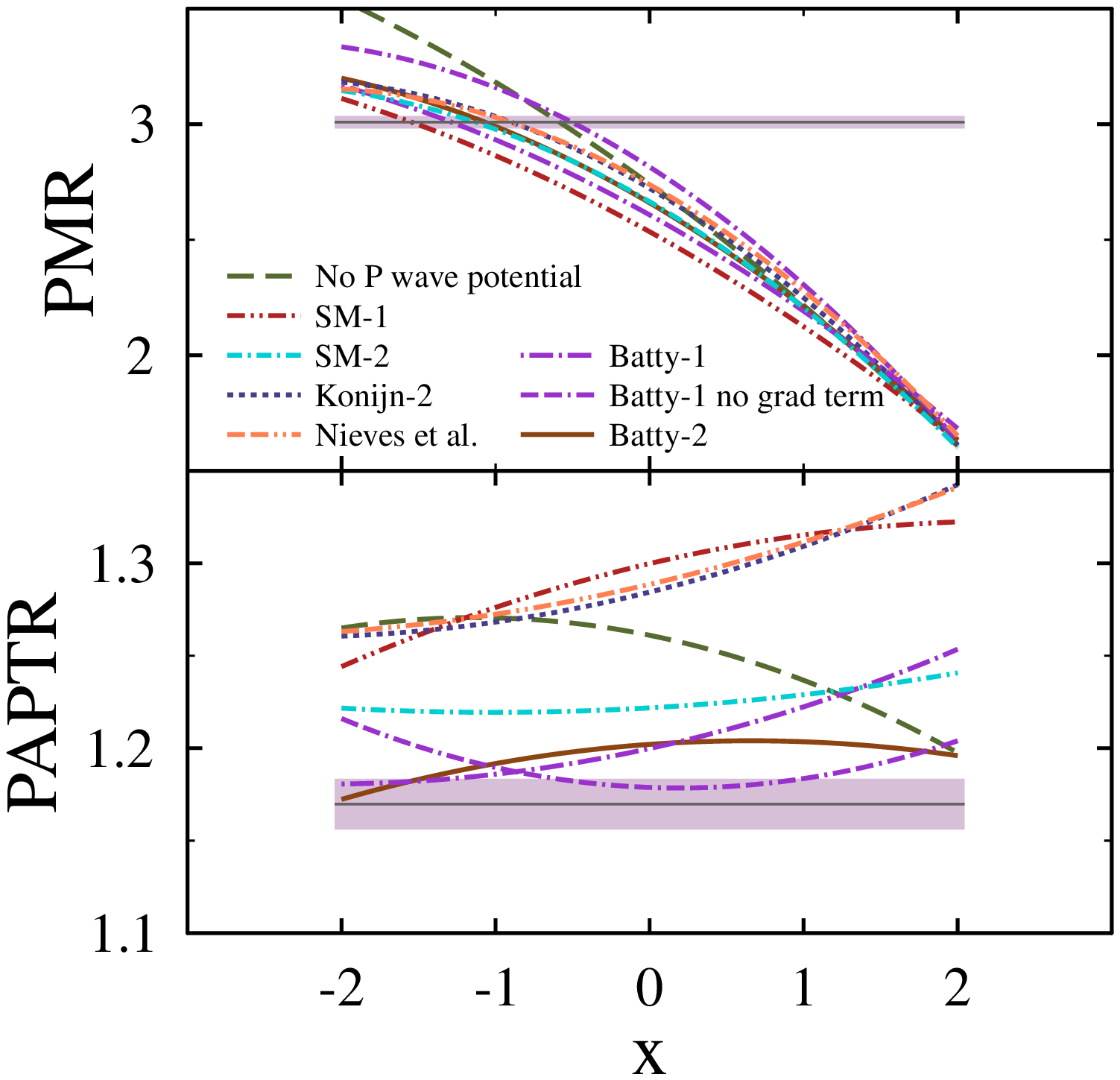}
\end{minipage}
\end{center}
\caption{\figlab{sppotimpactratios} (Color online)  Impact of the various S (left panel) and P (right panel)
 wave pion potentials introduced in \secref{pionoptpot} on pion multiplicity ratio (PMR) and pion average $p_T$ ratio (PAPTR)
in mid-central collisions (3.35 fm $<$ b $<$ 6.00 fm) of $^{197}$Au+$^{197}$Au at an incident energy of 400 MeV/nucleon.
The meaning of the curves labeled ``No S wave potential'' and ``No P wave potential'' is the same as in~\figref{swavepotimpact}
and \figref{pwavepotimpact} respectively.}
\end{figure*}

The impact of the $S$-wave potential on pion multiplicities and average transverse momenta is presented in~\figref{swavepotimpact}. The case when contributions of the pion potential are omitted is presented as
a benchmark. The impact on multiplicities (left-hand panel) is stronger for the case of $\pi^-$ and generally a $S$-wave potential that is stronger
at saturation densities and above leads to a stronger decrease of the multiplicity in question. The most clearly visible example is
that of the Konijn-2 potential. The  weaker SM-1 and Nieves ${\it et\phantom{a}al.}$ potentials lead to the smallest change with respect to
the no pion potential case. Additionally, accounting for the empirical momentum dependence of the $S$-wave isoscalar strength
is also observed to impact only slightly pion multiplicities. Generally, the experimental values of charged pion multiplicities are
underestimated by the model by fractions that show a weak isospin dependence.

The right-hand panel of~\figref{swavepotimpact} presents the conclusions on the impact of the $S$-wave potential on the average transverse
momenta of charged pions. The cases of $\pi^-$ and $\pi^+$ mesons are clearly different. In the case of negatively charged pions
the impact always leads to higher average transverse momenta, the strength of the modification being, similarly to pion multiplicities,
in close correlation to the strength of the potential close to saturation density. Inclusion of $S$-wave potential leads predominantly to
values that overestimate the experimental FOPI value by at most 10$\%$. In the case of the positively charged pions the sign of the 
effect of the $S$-wave potential on average $p_T$ varies with the chosen potential, but in all cases the experimental values are over-predicted
by amounts in the range 5-20$\%$. For the case of the theoretical potential of Nieves ${\it et\phantom{a}al.}$ the impact is again among
the smallest for both charged pion states. The impact of the momentum dependence part of the potential on transverse momenta is 
however clearly visible, leading to increases of these observables by 5-10$\%$.

The impact of the $P$-wave potential is presented in~\figref{pwavepotimpact}. In all cases presented
in this figure the momentum dependent effective $S$-wave potential describe in~\secref{pionoptpot} and summarized in the paragraph
adjacent to~\eqref{b0b1final} has also been included, allowing a comparison of the full model with the experimental data. 
From the left-hand panel of this figure the impact of the $P$-wave potential on multiplicities can be inferred. For both charged states it stays
below 10$\%$, but is in relative magnitude bigger for the positively charged pion. The impact of the density gradient term of the potential
is visible particularly for the $\pi^+$ meson leading to an increase of a few percent of its multiplicity.

The impact of the $P$-wave pion potential on average $p_T$ values is clearly more important as can be seen from the right-hand panel 
of~\figref{pwavepotimpact}. Its attractive nature leads to lower values of $p_T$ for all presented choices for the $P$-wave potential,
the relative impact amounting to as much as 15$\%$. Generally, the experimental values of the $\pi^-$ and $\pi^+$ transverse
momenta cannot be described simultaneously. This suggests that the isovector part of the pion potential, as included in the present
model, is not accurate enough, either in strength or density dependence. Additionally the density gradient term of the $P$-wave potential is
 seen to have a discernible effect, at a few percent level, for both shown charge states of the $\pi$ meson. 
This result leads one to speculate on the possible 
relevance of the omitted isospin asymmetry gradient term in the $P$-wave potential.

It is noteworthy to investigate separately the impact of pion potentials on the
observables of primary interest for constraining the density dependence of the symmetry energy.  
To this end the influence of the $S$- and $P$-wave potentials on PMR and PAPTR in mid-central collisions of
$^{197}$Au+$^{197}$Au nuclei at 400 MeV/nucleon impact energy are presented in the left-hand and 
right-hand panels of~\figref{sppotimpactratios}, respectively. They are obviously derived
from the information presented in previous figures.

The inclusion of the $S$-wave pion potential leads in the majority of cases to a smaller PMR, the impact
on the extracted slope parameter $L$ of the SE amounting to as much as 20 MeV towards stiffer values. 
The momentum dependence of its isoscalar component influences the PMR only modestly. A similar
conclusion holds also for the $P$-wave pion potential impact, the value of the PMR is further reduced, pushing
the extracted stiffness of the asy-EoS to even higher values. From the right-hand panel of~\figref{sppotimpactratios}
an estimated impact of 20-40 MeV is obtained. It is noteworthy to point out that the influence of the density
gradient term is important, pushing the PMR to lower values. Most of the impact of the $P$-wave pion potential
on PMR is due to its gradient term. The combined effect of the $S$- and $P$-wave pion potential is to lower the PMR and
consequently push the extracted values of $L$ towards higher values by as much as 40-60 MeV. This margin is comparable
to the precision achieved in constraining the slope of the symmetry energy at saturation using elliptical flow 
data of the FOPI-LAND collaboration~\cite{Russotto:2011hq,Wang:2014rva,Cozma:2013sja}, but a factor
2-3 larger than the foreseeable accuracy that will be reported in the near future using the experimental results 
for the same observable measured by the ASYEOS Collaboration~\cite{Russotto:2016ucm}.

Turning to PAPTR, it is readily observed that the inclusion of the $S$-wave pion potential leads, for all presented
choices of the potential, to smaller values of this observable. The model  generally leads to values of this observable
higher than the experimental one, the discrepancy growing larger towards stiffer values of $L$. As in the case of the PMR,
the impact of the momentum dependent part of the isoscalar part of the $S$-wave pion potential is small. The sign
of the contribution of the $P$-wave potentials to the final value of the PAPTR varies with the chosen potential. Only a
few of the $P$-wave potentials used in this study were able to lead to values of the PAPTR in agreement or close to
its experimental one. It should however be noted that for the study of impact of the $P$-wave potential the included
$S$-wave component was that of the so called ``full effective model'' which can be observed, from left panel of
~\figref{sppotimpactratios}, to lead to higher values of PAPTR than some $S$-wave potentials extracted 
from pionic atom data. The impact of the density gradient term of the $P$-wave potentials is at the level of a few
percent and leads to lower and higher value of PAPTR for the a stiff and soft asy-EoS, respectively. Furthermore
the direction of the impact (increase vs. decrease) was revealed to be, during this investigation, dependent 
also on the choice of the $P$-wave potential. It can be concluded that for an accurate description of the experimental 
value of PAPTR both the $S$- and $P$-wave pion potentials need to be precisely known from other sources if models that
take into consideration particle production threshold effects and enforce the conservation of the total energy are employed.

A comparison of predictions of the full model with experimental rapidity and transverse momentum spectra of 
pions~\cite{Reisdorf:2015aa} will be postponed until the end of next section, in order to be able to make use of the
extracted strength, from experimental data, of the isovector $\Delta$(1232) potential.

\begin{figure}
 \includegraphics[width=16.0pc]{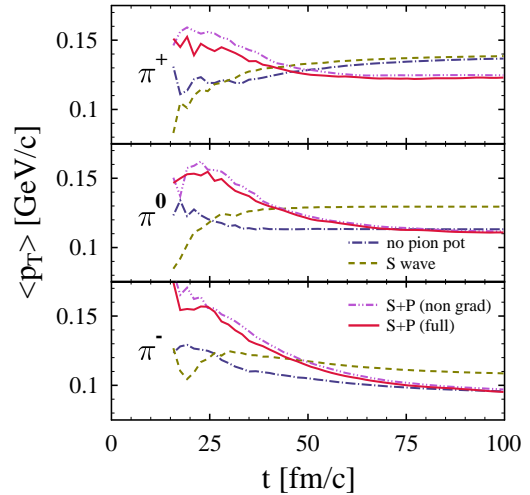}
\caption{\figlab{timedepavpt}(Color online) Time dependence of the average transverse momentum of
$\pi^-$ (top), $\pi^0$ (middle) and $\pi^+$ (bottom) mesons in mid-central heavy-ion collisions of 
$^{197}$Au+$^{197}$Au at an incident energy of 400 MeV/nucleon for several choices for the total potential
experienced by pions besides Coulomb. The value of the stiffness parameter has been set to $x$=0 and the Batty-1 parametrization
of the $P$-wave pion potential has been selected where indicated. The labels bear the following meaning: ``no pion pot'', both
components of the pion potential have been omitted; ``S-wave'', only the  effective model $S$-wave pion potential
has been included; ''S+P(non grad)``, the $S$-wave and non-gradient terms of the $P$-wave potential have been included;
and finally ''S+P(full)``, the $S$-wave and the full $P$-wave (both the gradient and non-gradient terms) are taken into account.
The vast majority of pions that escape into detectors are emitted at moments ulterior to t=30 fm/c.}
\end{figure}

This section is concluded with a presentation, in~\figref{timedepavpt}, of the time evolution of the 
average transverse momenta of the three charge states of the pion during heavy-ion collisions. Four cases
have been selected for this study. For the first one the effect of pion potentials is completely neglected (labeled ``no pion pot'').
The other three are obtained by successively adding the following ingredients to the first case: the $S$-wave
pion potential (``S wave''), non density-gradient terms of the $P$-wave pion potential (``S+P (non grad)'') and the density gradient term
of the $P$-wave potential (``S+P(full''). Additions of each of these contribution leads to important modifications of the average $p_T$
of pions at earlier stages of the collision, particularly during the high density phase. 
The impact of the $S$- and $P$-wave components of the potential are of comparable magnitude, there is however
a noticeable isospin dependence for the former one. The influence of the gradient terms of the $P$-wave potentials
on the final values of transverse momenta is smallest, at a few percent level. 
It is however clear that the outcome of the full model is the result of fine (partial) cancellations of the effects of all
the components of the pion potential, making the need for their precise knowledge more transparent.

\section{Constraining the symmetry energy}
\seclab{secon}

Given the modifications of the transport model described in~\secref{transpmodel} and the inclusion
of the pion optical potential it is worthwhile to rediscuss the impact of the $\Delta$(1232) isovector
potential on the PMR and stress the differences with the conclusions of the initial investigation reported in Ref.
~\cite{Cozma:2014yna}. The analysis will be extended to include the PAPTR and also to a wider range of the
strength of the isovector $\Delta$(1232) potential. The latter quantity will be allowed both attractive and repulsive
strengths in the range [-2,3] in units of $\delta$ (see~\eqref{choicedeltapot2} and the paragraph following it).
The effects due to the effective model for the $S$-wave pion potential and Batty-1 parametrization for the $P$-wave one
(with an energy dependence as discussed in~\secref{pionoptpot}) have been included.
The results of this simulation are presented in~\figref{avptratdelpotdep}.

The impact of the strength of the isovector $\Delta$(1232) potential on the PMR is presented in the left-hand panel
of~\figref{avptratdelpotdep} for different values of the asy-EoS stiffness parameter $x$. The strong dependence
of the PMR on this quantity for all values of $x$, with the exception of a narrow interval that encloses $x=1$,
is evident. The sensitivity decreases however for attractive choices of the strength $V_v$ and  becomes rather
small in the neighborhood of $V_v=-2\delta$. As already noted at
the beginning of~\secref{pipotobs}, the small changes implemented to the transport model, with respect to 
Ref.~\cite{Cozma:2014yna}, lead to a slightly lower value for PMR and somewhat modified dependence on $x$, 
mostly due to the decrease of the strength of the Coulomb interaction. This is most easily visible for the
a strength of the isovector $\Delta$(1232) potential $V_v$=3$\delta$. For this case the current model leads 
to PMR values that are largely independent of $x$, while the version of Ref.~\cite{Cozma:2014yna} gives rise
to increasing PMRs with increasing SE softness allowing the description of the experimental
FOPI value for a very soft asy-EoS. Extrapolations of the presented results suggest that the current model
would be able to describe the experimental PMR data for a stronger than $V_v$=3$\delta$ isovector $\Delta$(1232)
potential and a soft SE, in addition to the cases evident from the left panel of~\figref{avptratdelpotdep}.

\begin{figure*}[htb]
\begin{center}
\begin{minipage}{0.49\textwidth}
\includegraphics[width=16.0pc]{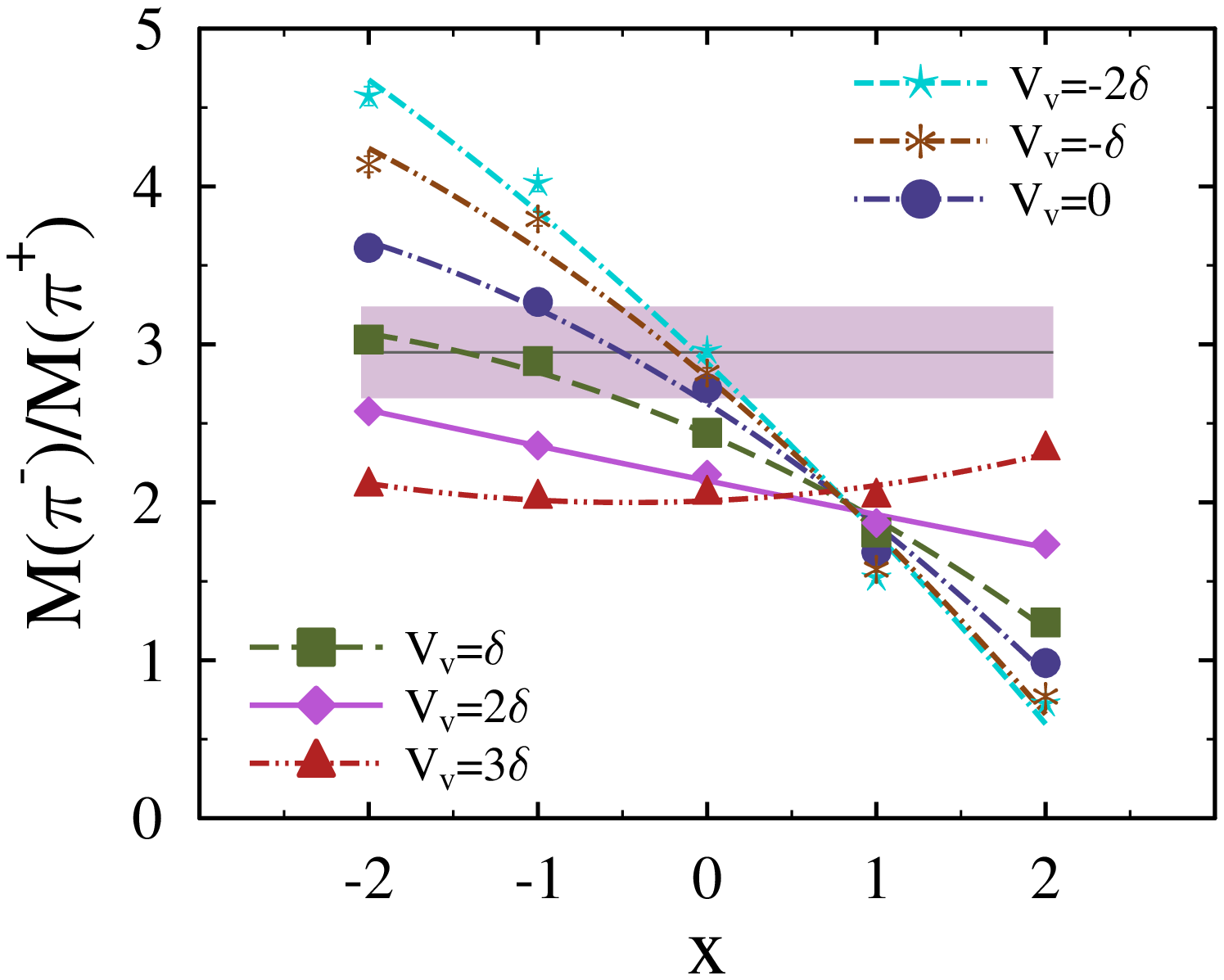}
\end{minipage}
\begin{minipage}{0.49\textwidth}
\includegraphics[width=16.75pc]{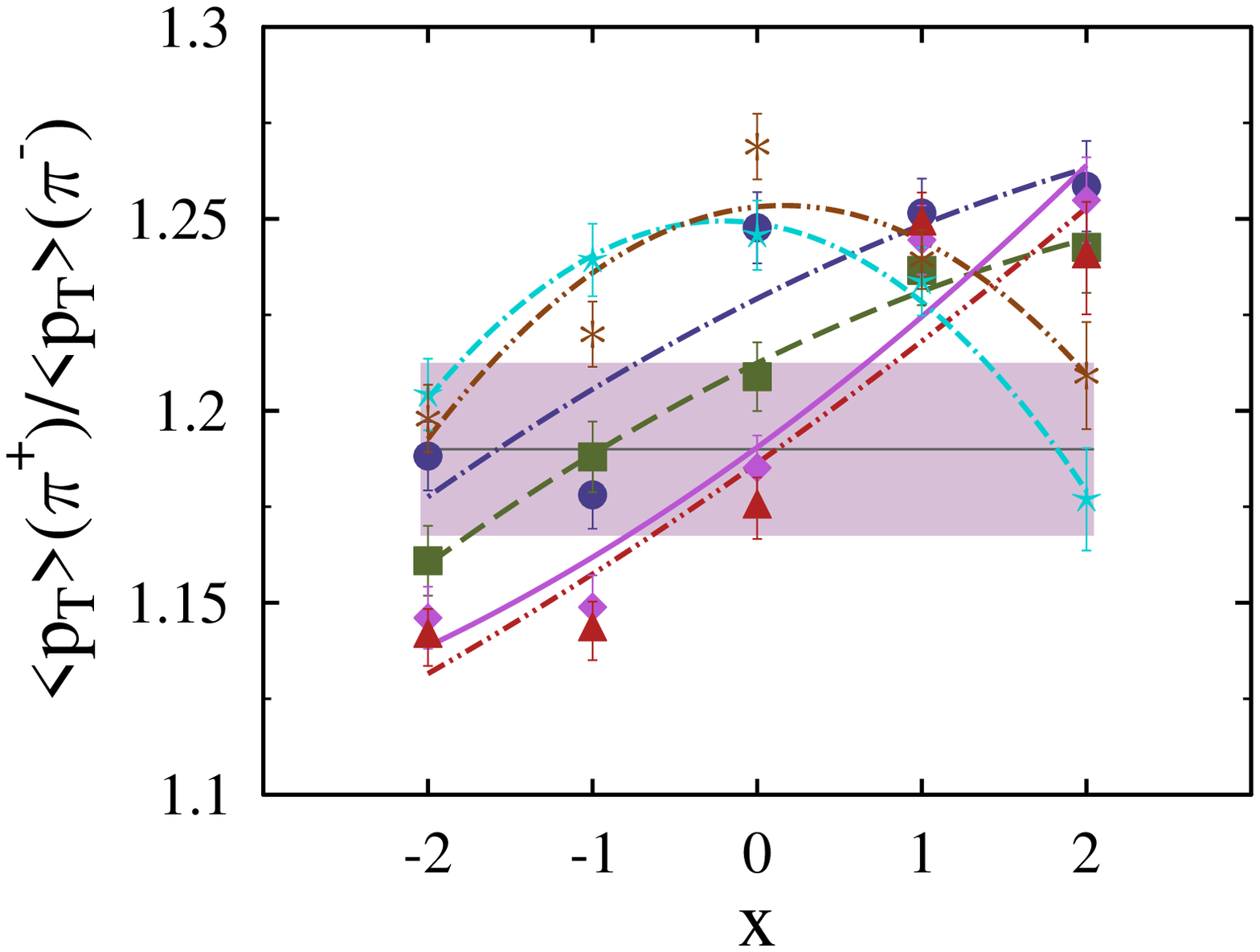}
\end{minipage}
\end{center}
\caption{\figlab{avptratdelpotdep}(Color online) The ratio of charged pion multiplicities (left panel) and
average transverse momenta (right panel) as a function of the stiffness parameter $x$ for six
different choices of the strength of the isovector component of the $\Delta$(1232) potential in nuclear
matter (see~\eqref{choicedeltapot2}). The results correspond to central (b$<$2.0 fm) $^{197}$Au+$^{197}$Au 
collisions. The full experimental FOPI results (including both systematical and statistical errors) 
of Refs.~\cite{Reisdorf:2006ie,Reisdorf:2010aa} are depicted by horizontal bands.}
\end{figure*}

\begin{figure*}[htb]
\begin{center}
\begin{minipage}{0.49\textwidth}
\includegraphics[width=16.0pc]{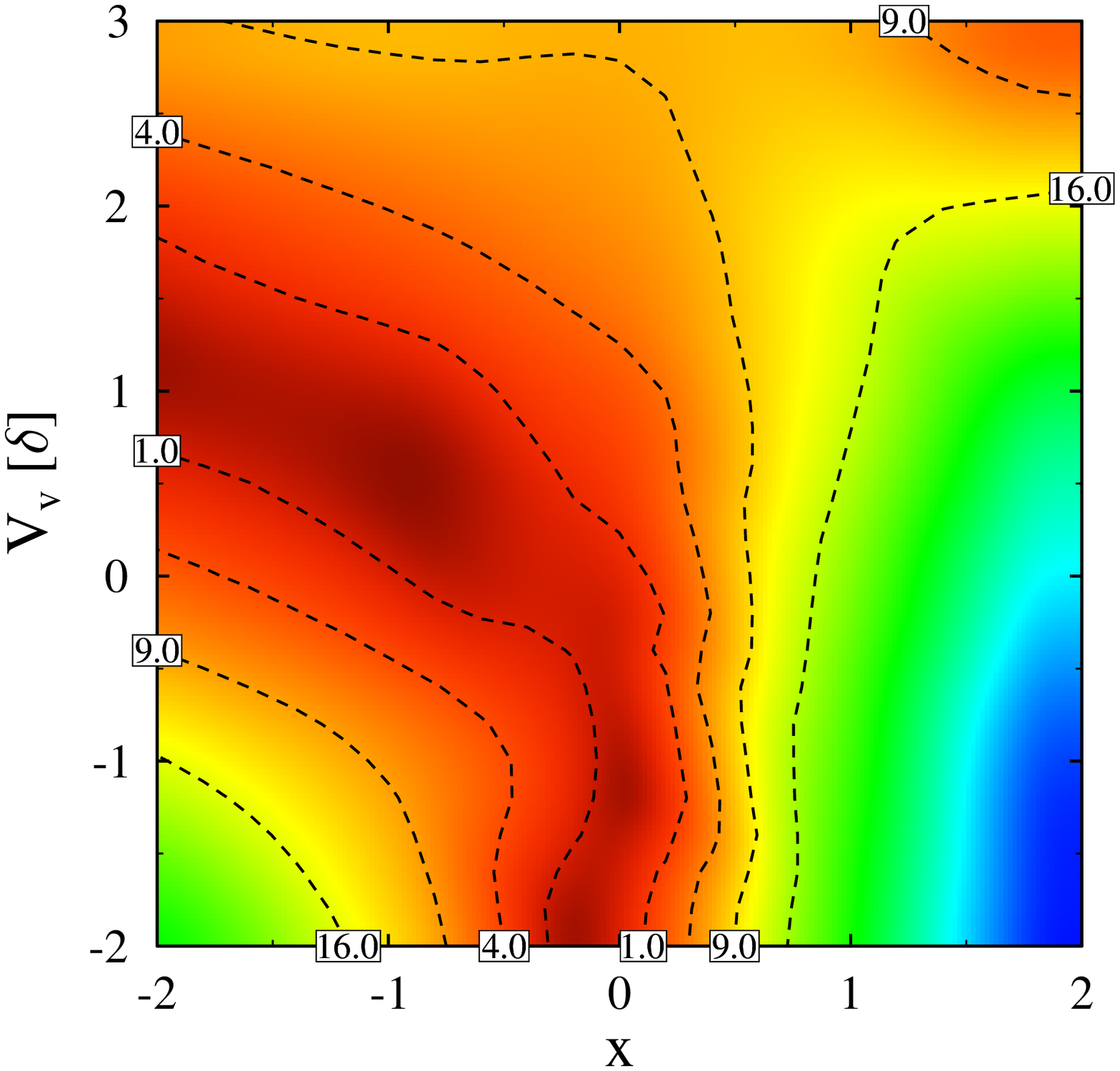}
\end{minipage}
\begin{minipage}{0.49\textwidth}
\includegraphics[width=16.0pc]{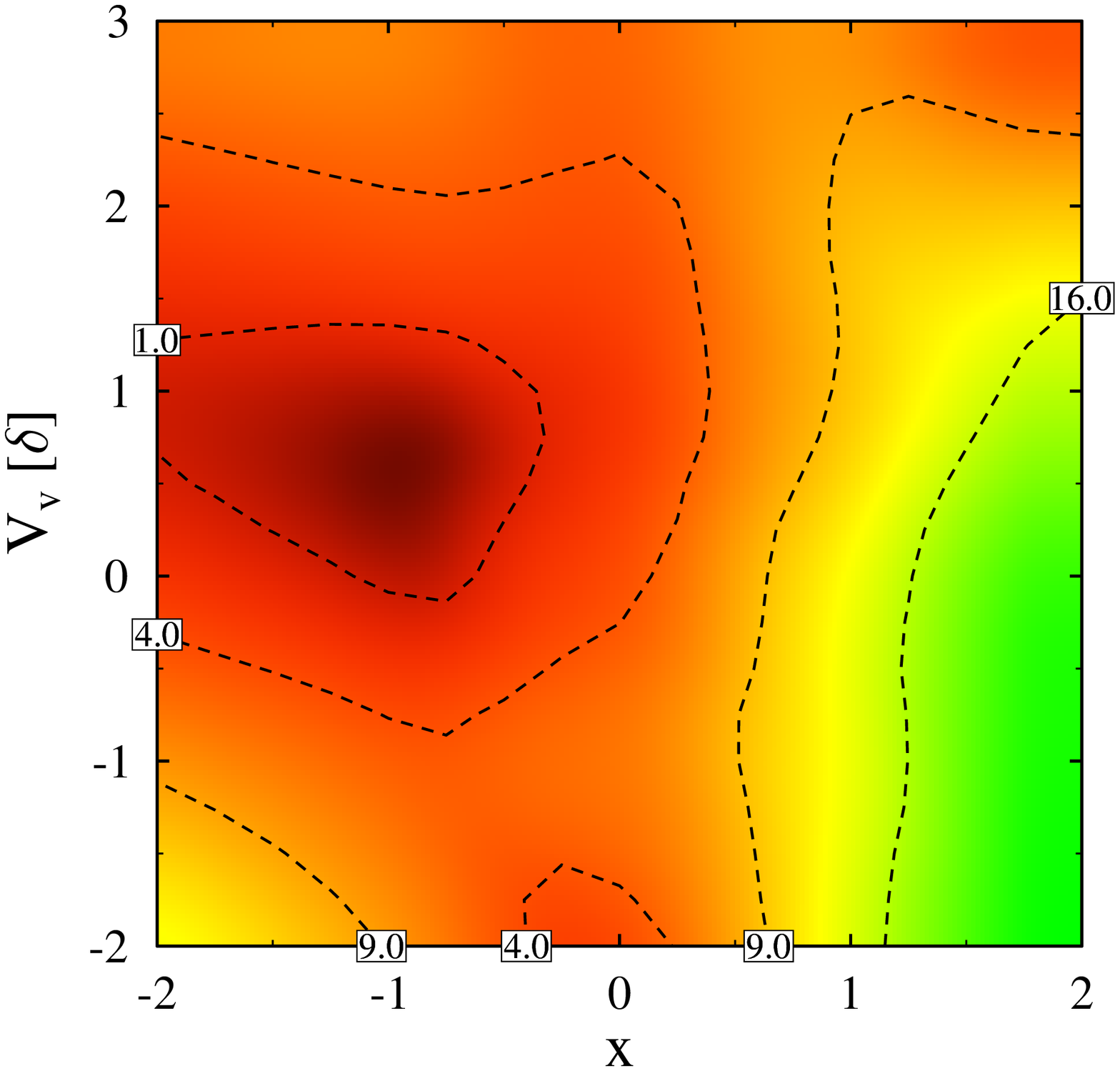}
\end{minipage}
\end{center}
\caption{\figlab{chisqdivsigbat1} (Color online) Contour plots for the $\chi^2/\mathrm{dof}$ value of the comparison
theory vs. experiment for the case of the PMR (left panel) and PMR+PAPTR (right panel). In the 
latter case, contributions due to the two observables are added with the same weights. Curves are labeled according
to the corresponding value of the $\chi^2/\mathrm{dof}$ quantity.}
\end{figure*}

The right-hand panel of~\figref{avptratdelpotdep} presents the dependence of the PAPTR on both the strength of
the isovector $\Delta$(1232) potential and the SE stiffness parameter $x$. By fixing the former it can be concluded
that the PAPTR is sensitive to the isovector part of the EoS. The sensitivity to its stiffness is however considerably
less pronounced than for PMRs, amounting to at most 10$\%$ between the very soft and very stiff choices for $x$. 
This feature is more clearly visible for repulsive values of $V_v$ and in particular
for the choice commonly employed in most transport models $V_v$=$\delta$. For attractive $V_v$ the sensitivity
to the SE decreases to about 5$\%$, but for these cases the theoretical PAPTR values tend to over-predict the experimental
one. With regard to the extraction of constraints for the SE stiffness it should be noted that experimental values for
this observable are determined with much higher accuracy than for PMR (2.5$\%$ versus 10$\%$) which balances to
a certain extent the disadvantage of a lower sensitivity to the asy-EoS stiffness.
The sensitivity of the PAPTR to $V_v$ mirrors almost perfectly the behavior evidenced for the PMR. 
It reaches a maximum for stiff choices of the asy-EoS and becomes smaller for soft ones, vanishing in the neighborhood of
$x=1$. 

It is instructive to present the comparison theory versus experiment as $\chi^2$/dof plots for the observables
of interest that also exhibit the above discussed sensitivity to the stiffness of the symmetry energy and
strength of the $\Delta$(1232) isovector potential. This is achieved in the left panel of~\figref{chisqdivsigbat1}
for the PMR. To facilitate the extraction of information about the favored value of $x$ (and $V_v$) curves
for the 68$\%$, 95.5$\%$, 99.3$\%$ and 99.994$\%$ confidence levels, that allow the determination of
1, 2, 3 and 4 $\sigma$ uncertainties on the extracted value of the desired parameter, are plotted. They are
labeled by the corresponding value of $\chi^2$/dof. The conclusions of Ref.~\cite{Cozma:2014yna}, which can
also be inferred from the left panel of~\figref{avptratdelpotdep}, with regard to suitability of the PMR for the extraction of 
constraints for the density dependence of SE above saturation are more transparent.
Specifically, the extracted value for $x$ depends strongly on $V_v$ and, furthermore, for a repulsive isovector
$\Delta$(1232) potential the uncertainty increases as a result of the lower sensitivity of PMR to the asy-EoS stiffness.
In contrast, for the hypothetical case of an attractive isovector $\Delta$(1232) potential the stiffness of the SE can be more accurately
determined and is almost independent of the value of $V_v$.

\begin{figure*}[htb]
\begin{center}
\begin{minipage}{0.49\textwidth}
\includegraphics[width=17.0pc]{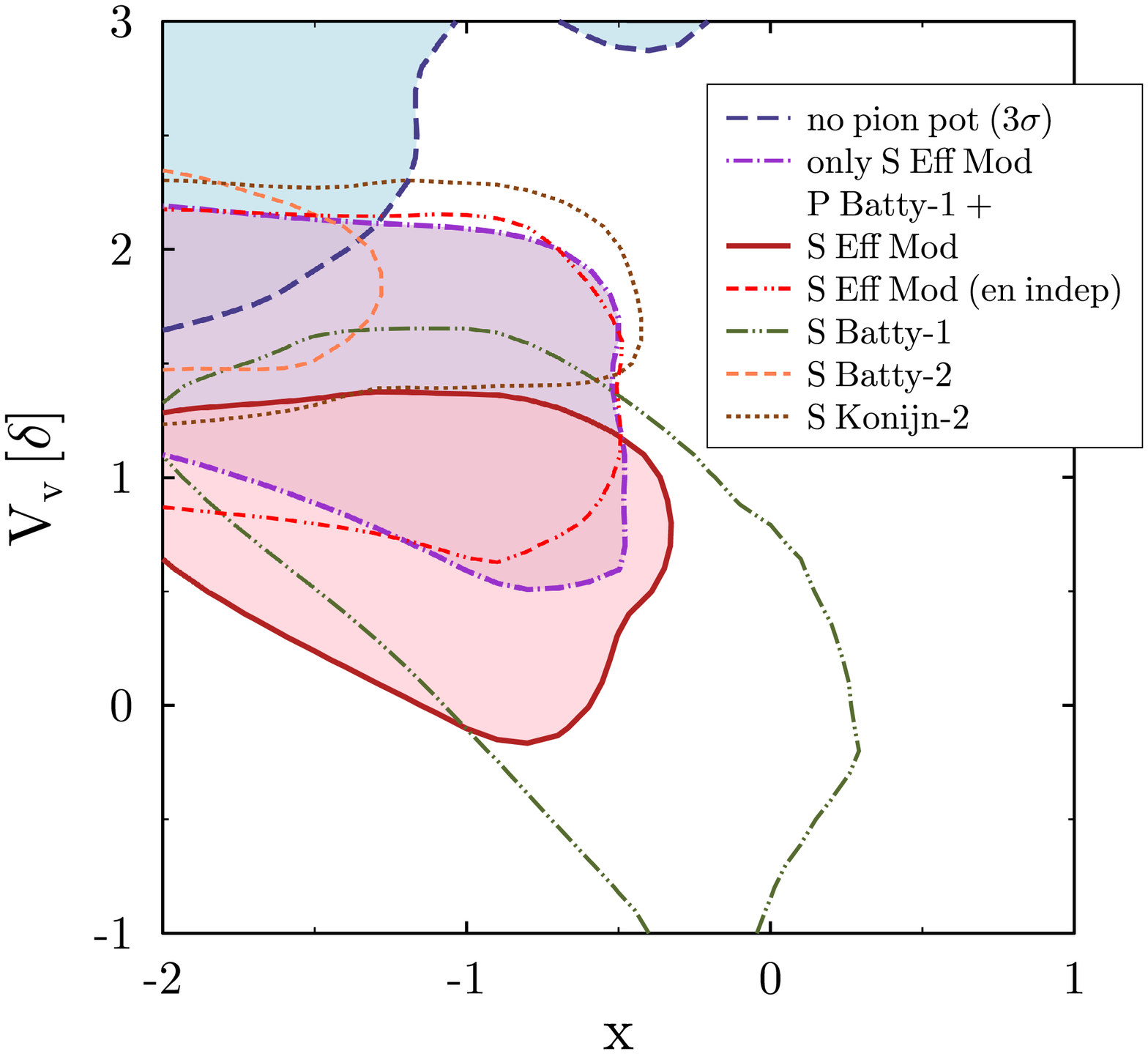}  
\end{minipage}
\begin{minipage}{0.49\textwidth}
\includegraphics[width=16.0pc]{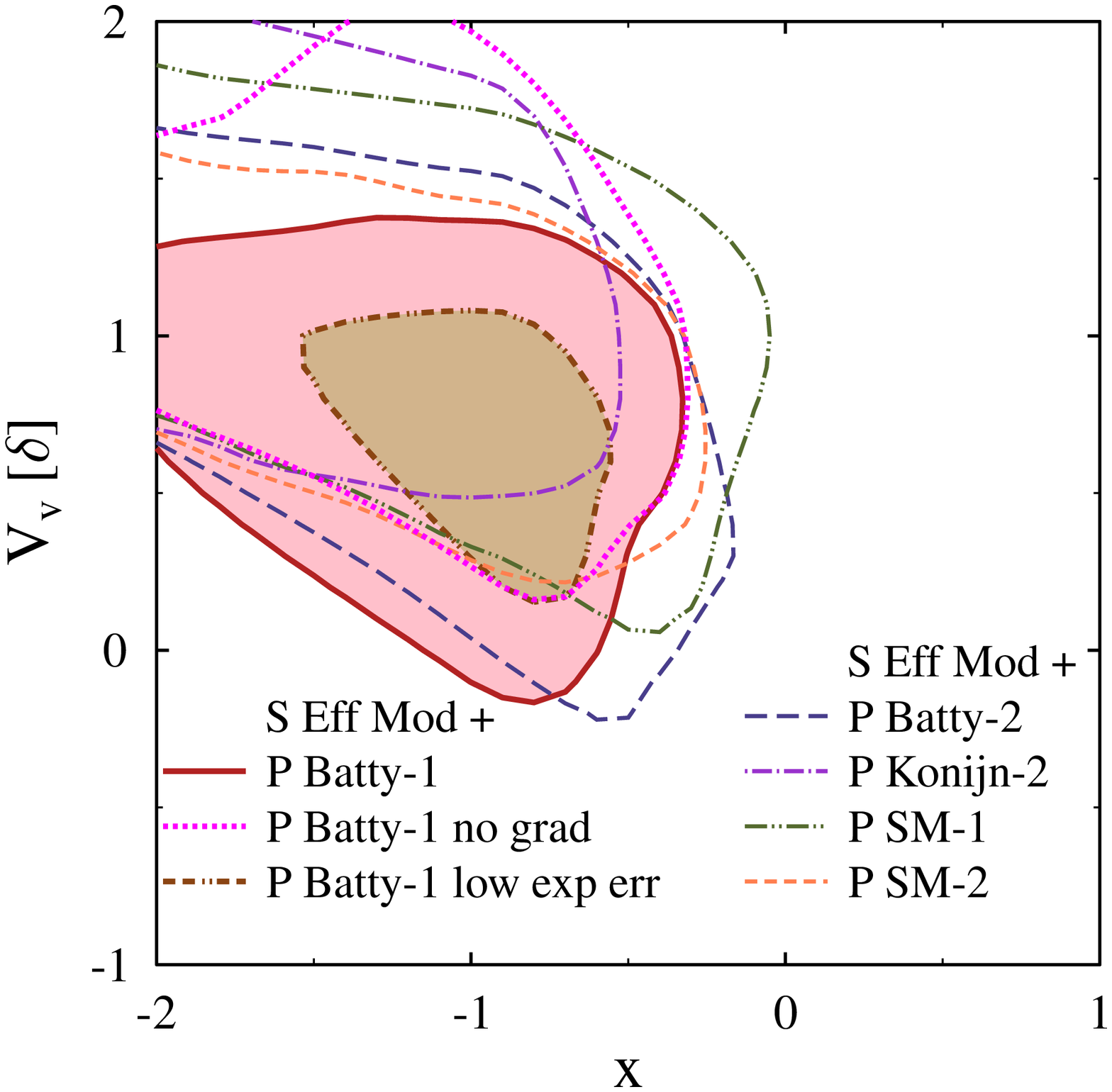} 
\end{minipage}
\end{center}
\caption{\figlab{chisq1sigdivpipot}(Color online) Sensitivity of the extracted constraints for the stiffness
parameter $x$ and strength of the isovector $\Delta$(1232) potential $V_v$ to different choices for the $S$-wave (left panel)
and $P$-wave (right panel) pion potentials. The calculations help quantify the impact of uncertainties in the energy and density dependence of
these potentials on the quantities of interest. For that purpose the 1$\sigma$ confidence level (CL) contour plots of the 
quantity $\chi^2/\mathrm{dof}$ determined by comparing theoretical and experimental results for the observables PMR and PAPTR for 
different choices for the pion $S$-wave (left panel) or $P$-wave (right panel) potentials are plotted. The case for which pion potential
contributions are completely omitted (''no pion pot``) or only the $S$-wave component is included (''only S Eff Mod``) are also shown
(left panel). Result for a version of the $P$-wave potential with the gradient terms omitted is also shown 
(``P Batty-1 no grad'' in the right panel).
Additionally, for the Batty-1 parametrization of the $P$-wave potential the 1$\sigma$ CL contour curve determined by artificially
decreasing the experimental uncertainties for PMR and PAPTR to 3$\%$ and 1.5$\%$, respectively, is also plotted and labeled
``P Batty-1 low exp err'' (right panel).}
\end{figure*}

The observed impass can be resolved by including in the expression of $\chi^2$/dof contributions due to
PAPTR. This claim is supported by the results presented in the right-hand plot of~\figref{chisqdivsigbat1}. 
It is observed that the range of allowed, at 1$\sigma$ confidence level, isovector $\Delta$(1232) potential
strength values $V_v$ is significantly narrower, favoring a mildly repulsive one, as compared to the case when only
contribution due to PMR are included. Similarly, the allowed range for the stiffness parameter $x$ is more
constrained at 1$\sigma$ level to $x$=-1.0$^{+0.75}_{-1.5}$. The corresponding value for the allowed slope parameter interval is
$L$=106$^{+67}_{-34}$ MeV. The accuracy is comparable with the one that can be achieved from elliptic flow ratio constraints
that make use of the FOPI-LAND experimental data~\cite{Russotto:2011hq,Cozma:2013sja,Wang:2014rva}, but is a factor
of 2-3 more imprecise than what can be accomplished by using the most recent ASYEOS Collaboration results for similar observables
~\cite{Russotto:2016ucm}. The less than optimal accuracy when extracting the value of $L$ from pion related observables 
originates from three main sources: first, the sensitivity of the PMR to the SE stiffness decreases towards higher values of $L$; 
second, the experimental uncertainty on PMR amounts to a rather large value, close to 10$\%$; third, the sensitivity of PAPTR
to the slope parameter $L$ is not as pronounced as for the PMR.

Each of these three sources of uncertainties can be reduced in the following manner: (1) choosing nuclei with higher isospin
asymmetry; (2) performing experimental measurements at lower impact energies, closer or even below the vacuum pion
production threshold and (3) improving experimental accuracy. All of these requirements will be fulfilled by measurements
that will be performed in the very near future by the SAMURAI TPC collaboration~\cite{Shane:2014tsa}. For the already
existing FOPI experimental data only the last source can be partially alleviated by performing a reanalysis of the available
data sets and excluding from the spectra the regions of increased systematic uncertainties, as is for example the low-energy
part of the pion spectrum.

Before such an improved experimental result will become available, it will be useful to attempt to estimate the impact of such
an effort on the extracted constraints for the slope parameter $L$ and isovector $\Delta$(1232) potential strength $V_v$.
This exercise will also offer indications about the potential of the experimental program put forward by the
SAMURAI TPC collaboration~\cite{Shane:2014tsa}. To this end, the uncertainties of the experimental
FOPI values for the PMR and PAPTR in central $^{197}$Au+$^{197}$Au collisions at an impact energy of 400 MeV/nucleon 
have been decreased artificially from
9.5$\%$ to 3$\%$ and from 2.5$\%$ to 1.5$\%$ respectively. The results are plotted in the right-hand panel of
~\figref{chisq1sigdivpipot} for the case
when Batty-1 $P$-wave pion potential is used by the double-dashed-double-dotted curve (labeled ``Batty-1 low exp err''). The case when
the full magnitude of the experimental uncertainties is considered is depicted by the full curve (labeled ``Batty-1''). The
important ``improvement'' of the experimental accuracy leads to an increase of the accuracy of the extracted values
for $L$ and $V_v$ by fraction amounting to about 30-40$\%$ of the old result. It can thus be concluded that a precise determination
of the slope parameter from pion observables will require a very careful choice of the system studied and of the impact energy together
with a significant improvement of the experimental accuracy to values in the few percent range.

The success of such a program will only be warranted if certain progress on the theoretical side, mainly a better knowledge
of the pion potential away from the density and kinetic energy region that was constrained using pionic atoms and pion-nucleus scattering,
will be also achieved. To support this statement, the 1$\sigma$ confidence level (CL), if not otherwise specified,
of the theoretical versus experimental comparison of PMR+PAPTR has been plotted in ~\figref{chisq1sigdivpipot}
for various choices of the pion $S$-wave (left panel) and $P$-wave (right panel)
potential, while keeping the other component ($P$-wave for the left panel and $S$-wave for the right panel) the same. Such calculations
help quantify the model dependence introduced by extrapolating the pion potential far outside the density/momentum region probed
in pionic atom and pion-nucleus scattering experiments. 

\begin{figure*}[htb]
\begin{center}
\begin{minipage}{0.49\textwidth}
\includegraphics[width=16.5pc]{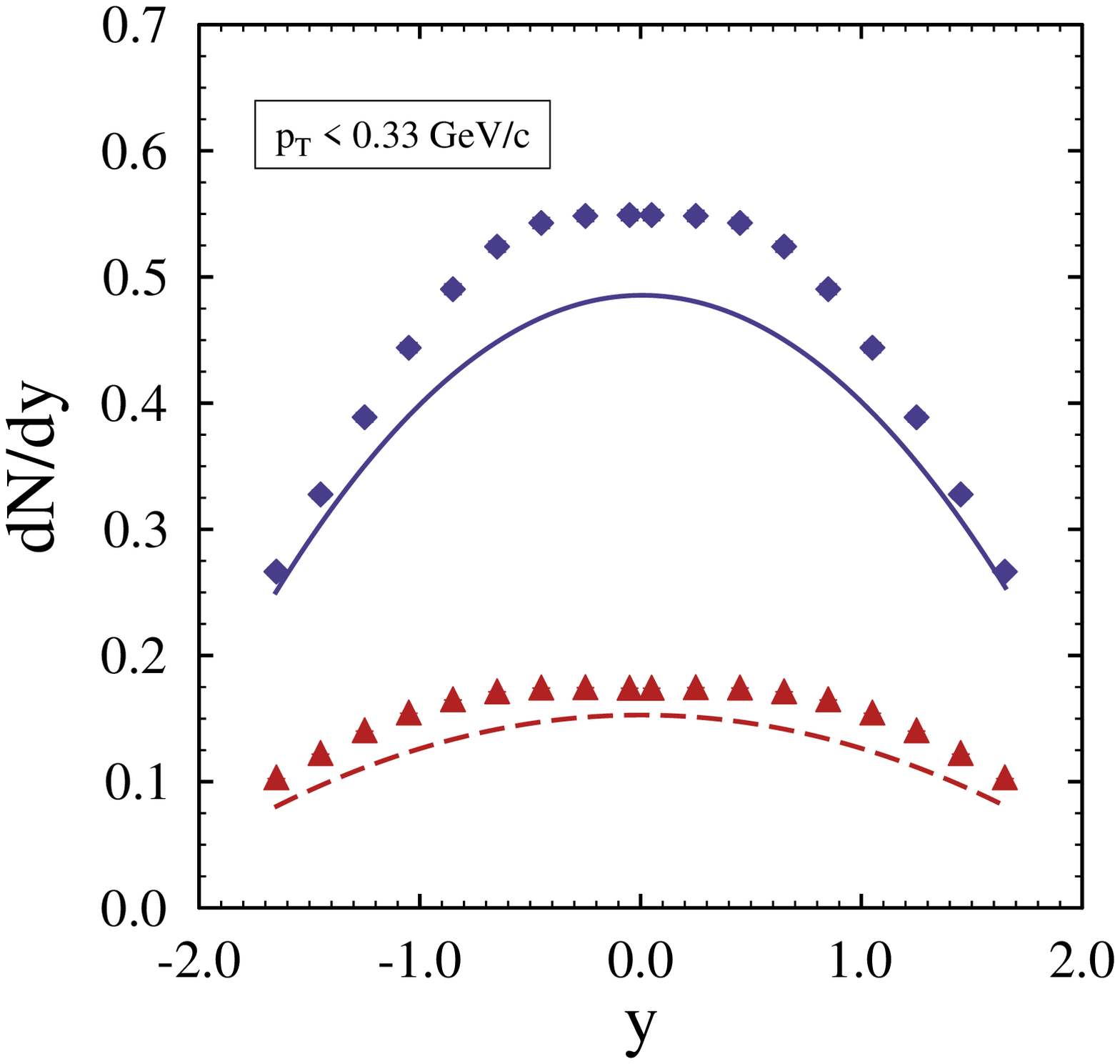}
\end{minipage}
\begin{minipage}{0.49\textwidth}
\includegraphics[width=16.0pc]{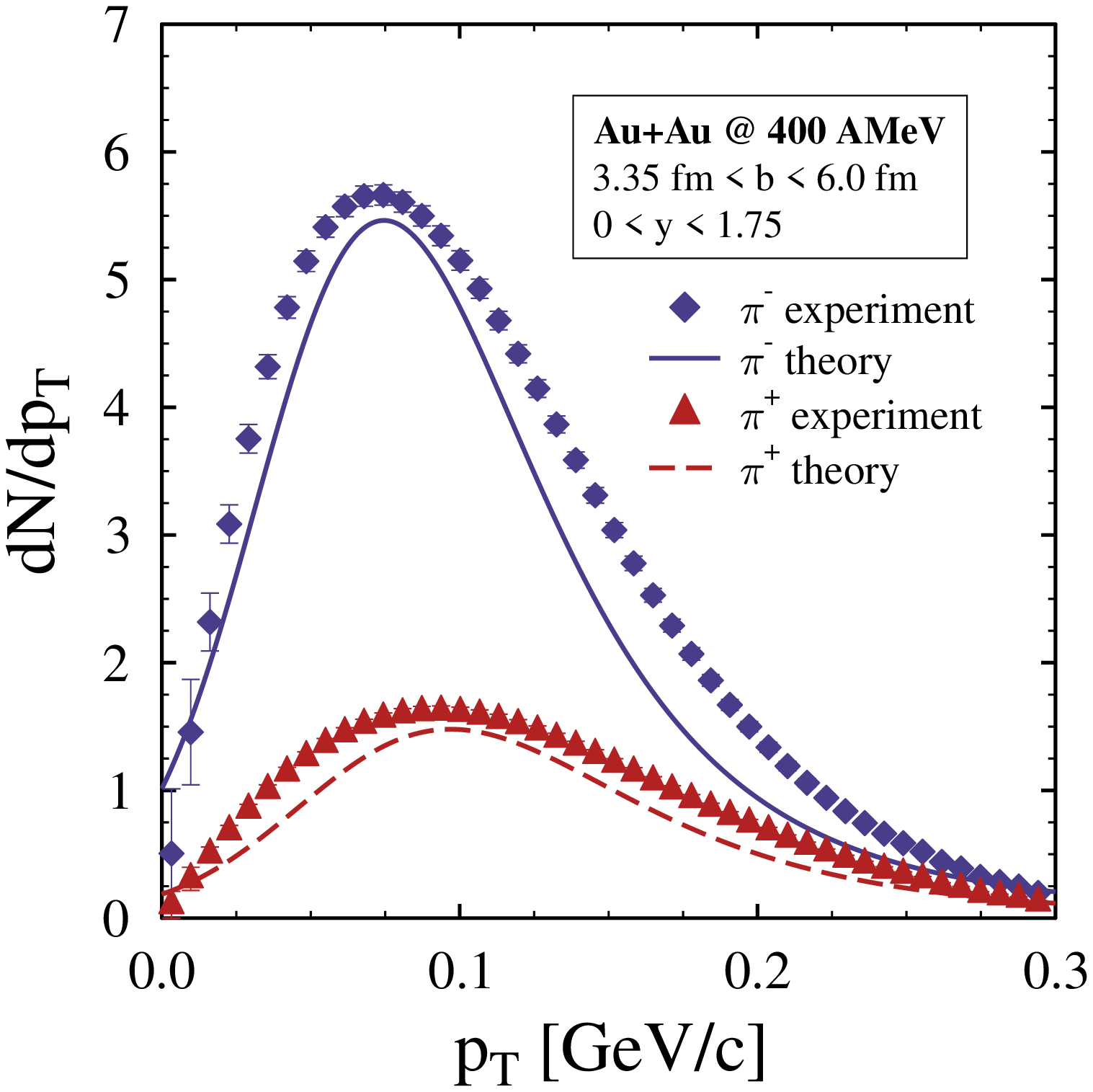}
\end{minipage}
\end{center}
\caption{\figlab{rapptspectra} (Color online) Rapidity (left-hand panel) and transverse momentum (right-hand panel) spectra
of charged pions in mid-central $^{197}$Au+$^{197}$Au collisions at impact energy of 400 MeV/nucleon.
Theoretical results for $\pi^-$ (full curves) and $\pi^+$ (dashed curves) mesons are compared with their
experimental counterpart~\cite{Reisdorf:2015aa} depicted by diamond and triangle shaped symbols respectively.
In addition to the impact parameter selection (3.35 fm$<$b$<$6.0 fm) the indicated kinematical cuts were imposed
in each case. The value of the stiffness parameter has been set to $x$=-1. Systematic uncertainties were not accounted for in the plotted experimental error bars.}
\end{figure*}

A calculation in which both pion potential components have been omitted has
also performed and it demonstrates that a simultaneous description of the experimental values for PMR and PAPTR cannot be achieved for
any reasonable choices of the $x$ and $V_v$ parameter values. In fact, most of the probed parameter space lies outside the 3$\sigma$ CL
region (see the dashed curve and the filled region enclosed by it in the left panel of~\figref{chisq1sigdivpipot}). 
The inclusion of the $S$-wave potential, with the $P$-wave potential switched off (``only S Eff Mod''), drastically improves the situation, 
the 1$\sigma$ CL region being almost entirely inside the parameter search window. Switching on the $P$-wave potential (``P Batty-1 + S Eff Mod'')
impacts visibly only the favored value of $V_v$, the extracted value for the stiffness parameter $x$ being virtually the same. A similar
conclusion holds true in regard to the impact of the energy dependent piece of the $S$-wave potential (compare double-dashed double dotted and
full curves in the left panel of~\figref{chisq1sigdivpipot}). Different density dependencies for both the isoscalar and isovector parts
of the $S$-wave pion potential do however have an important impact on the extracted values of both $x$ and $V_v$ parameters. The uncertainty
of the favored value for the slope $L$ can amount to as much as 60 MeV. In this context it should be recalled that all the employed $S$-wave potentials
have very similar strengths in the 0.5-0.75$\rho_0$ density region, but they differ significantly above saturation in either the isoscalar or the isovector
channel (see left panel of~\figref{swavepot}).

The magnitude of the model dependence due to uncertainties in the $P$-wave pion potential can be inferred from the right panel of ~\figref{chisq1sigdivpipot}.
Results for each of the parametrizations listed in~\tabref{pionpotparameters} are provided. It can be seen that the minimum allowed value for $L$ varies rather strongly
with the choice of the potential between the limits $L$=61 MeV ($x$=0.0) and $L$=85 MeV ($x$=-0.5). Furthermore, for
all of the used $P$-wave potentials the maximum value of $L$ satisfies the constraint $L>$150 MeV. Additionally, the impact of the gradient
terms of the $P$-wave potential can be assessed by comparing the full and dotted curves in the same plot. It is seen that they affect visibly
only the extracted value of the $V_v$ parameter.

Taking into account all evidenced sources of uncertainty one can deduce, from a comparison of the present model and the
available experimental FOPI data for PMR and PAPTR, that at 68$\%$ CL the slope of the symmetry energy at saturation
has to be stiffer than $L$=50 MeV irrespective of the strength of the $\Delta$(1232) potential or of details of the pion optical
potential. This lower limit is further decreased to roughly $L$=30 MeV and $L$=15 MeV to achieve a 
95.5$\%$ and 99.3$\%$ CL result. It can be therefore concluded, with sufficient certainty, that the density dependence of 
the symmetry energy is not soft. The extraction of an upper limit with a similar confidence level will have to be postponed 
until more precise data of a reaction more sensitive to the density dependence of the SE become available.

The present transport model favors, on average, when the Batty-1 parametrization for the $P$-wave potential is employed a rather
soft repulsive $\Delta$(1232) potential, with $V_v$=0.5$\delta$. Using this result, the charged pion rapidity and
transverse momentum spectra have been determined and are plotted in~\figref{rapptspectra} together with the FOPI
experimental result~\cite{Reisdorf:2015aa}. For this theory versus experiment comparison the value of the stiffness parameter has
been set to $x$=-1, which is close to the average value favored by the model in conjunction with the Batty-1 $P$-wave pion 
potential (see~\figref{chisq1sigdivpipot}). The comparison is performed for mid-central collisions of $^{197}$Au+$^{197}$Au
nuclei, the additional imposed kinematical constraints being 0$<y/y_P<$1.75 and $p_T<$0.33 GeV/c. It is observed that
both the rapidity and $p_T$ experimental spectra are under-predicted by the model by fractions in the range of 10$\%$-20$\%$
particularly in the mid-rapidity and higher than average $p_T$ regions. The high rapidity and high $p_T$ ends of
the shown spectra are generally in close agreement to the data. As the $\pi^-$ and $\pi^+$ multiplicities
increase and respectively decrease with the increase of asy-EoS stiffness the observed general under-prediction of experimental
data for these two observable may be reduced by a fine adjustment of the isoscalar part of the mean-field potential, in particular
that of the rather uncertainly known isoscalar $\Delta$(1232) potential. The momentum dependence of the difference theory versus
experiment underlines the need for a precise knowledge of the momentum dependent part of the optical potentials
the model relies upon.

\begin{figure}
 \includegraphics[width=16.0pc]{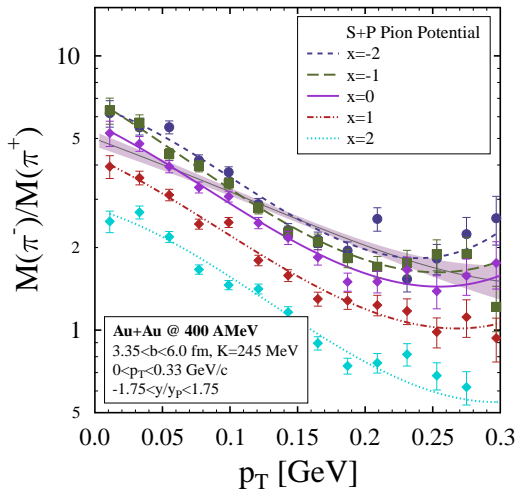}
\caption{\figlab{pmrptdep}(Color online) Theoretical transverse momentum PMR spectra in mid-central $^{197}$Au+$^{197}$Au 
collisions at an impact energy of 400 MeV/nucleon compared to its FOPI Collaboration experimental value~\cite{Reisdorf:2015aa}.
The effective model $S$-wave and the Batty-1 $P$-wave pion potentials have been accounted for in the simulations. Results for five
values of the stiffness parameter $x$ are plotted. Remarks made in the caption of~\figref{rapptspectra} holds also for 
the case presented here.}
\end{figure}

The same experimental data set can be used to perform a comparison theory versus experiment for the transverse momentum
dependent PMR. The results are presented in~\figref{pmrptdep} for all five choices of the stiffness parameter $x$
used in this study. The general feature of all the theoretical sets is that the slope of the transverse momentum dependent PMR is 
mildly stiffer than the experimental one at low and average values of $p_T$ and moderately softer at the high end part. 
As will become evident from~\figref{ptspectrapipotdep}
this observable is greatly influenced by the $S$-wave pion potential and to a lesser extent also by the $P$-wave one. An accurate
knowledge of the momentum dependent parts of the two components of the pion optical potential is therefore mandatory
for a successful description of the $p_T$ dependent PMR. It should be recalled that in this respect the approach employed in this study
has been rather qualitative, due to the lack of needed knowledge in this area. 
For the momentum dependent part of the $S$-wave pion potential a reproduction of the empirical values
of the effective isoscalar scattering amplitude $\bar{b}_0^{eff}$, which have been extracted from pion-nucleus scattering experiments,
have been imposed. 
In the case of the $P$-wave potential, the strength extracted from pionic atom data has been extrapolated to the energy of 
interest by mirroring the momentum dependence of the theoretical potential of Ref.~\cite{GarciaRecio:1989xa}. 
It is worth recalling that the experimental data plotted in~\figref{pmrptdep} do not include systematical uncertainties and
consequently the mild to moderate differences between the experimental and theoretical slopes of the $p_T$ dependent PMR
may not be statistically significant. Besides these observations, it is evident that the experimental data are qualitatively compatible 
with values of the stiffness parameter $x$=0 and $x$=-1 and to a lesser extent $x$=-2. On the other hand, soft choices 
of the asy-EoS stiffness, $x$=1 and $x$=2, under-predict the experimental values and will continue to do so also when the
systematical uncertainties of the experimental data, which are of the order of 10$\%$, will be taken into account.

\begin{figure*}[htb]
\begin{center}
\begin{minipage}{0.49\textwidth}
\includegraphics[width=16.0pc]{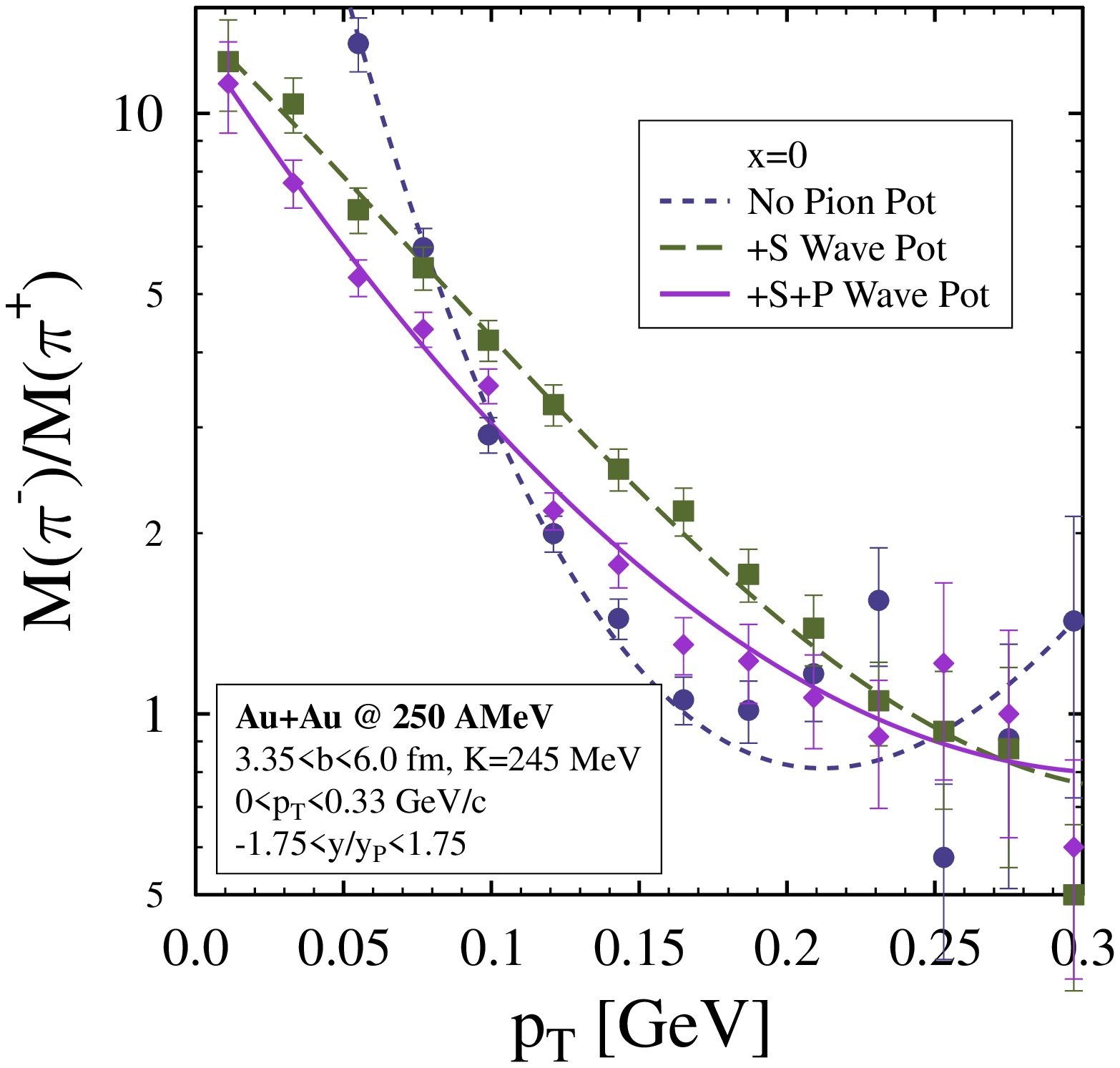}
\end{minipage}
\begin{minipage}{0.49\textwidth}
\includegraphics[width=16.0pc]{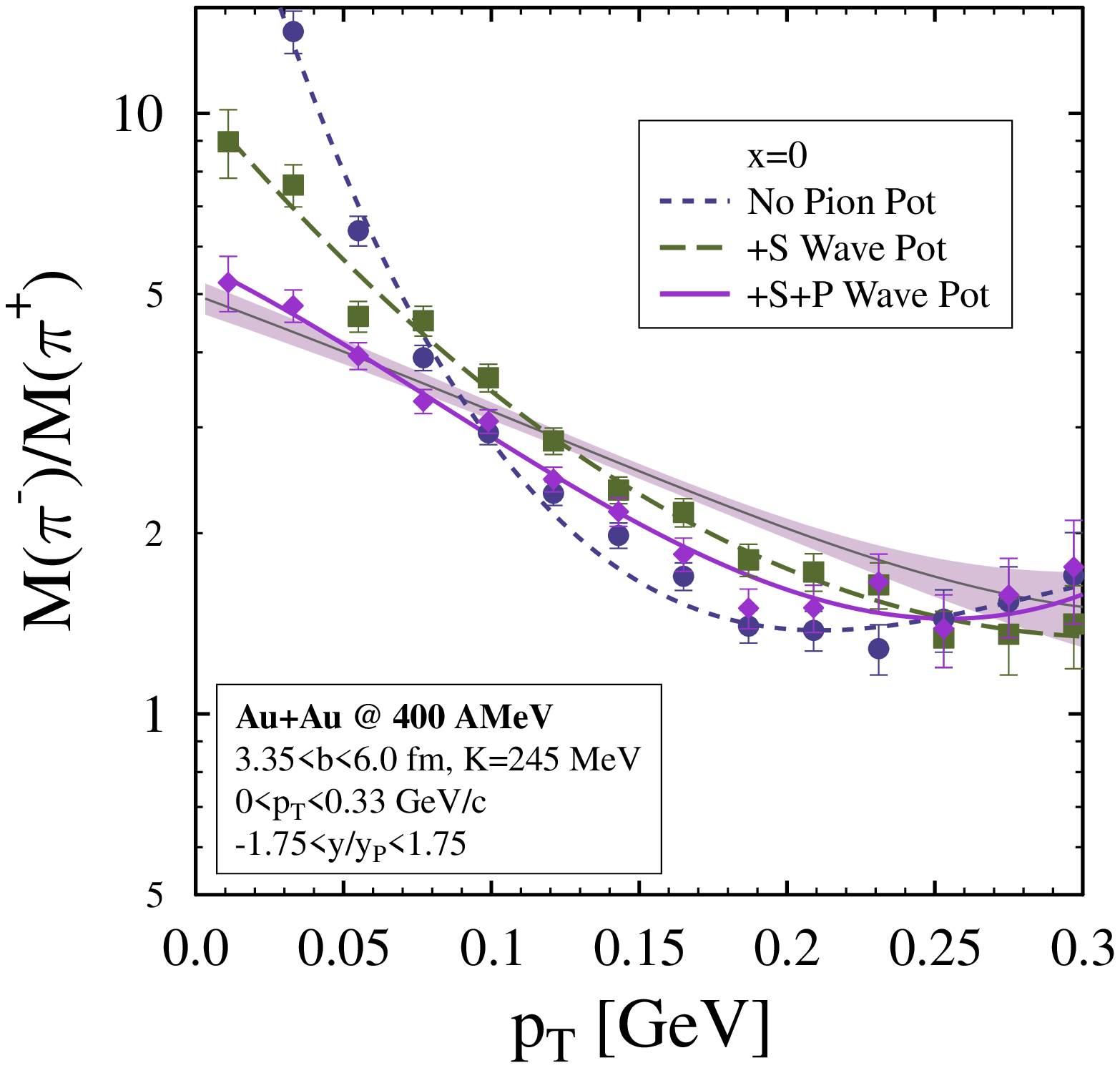}
\end{minipage}
\end{center}
\caption{\figlab{ptspectrapipotdep} (Color online) The impact of the $S$-wave and $P$-wave components of the
optical pion potentials on the transverse momentum PMR in mid-central $^{197}$Au+$^{197}$Au 
collisions for two different values of the impact energy: $T_{lab}$=250 MeV/nucleon (left panel) and
$T_{lab}$=400 MeV/nucleon (right panel). For the latter case the experimental FOPI result~\cite{Reisdorf:2015aa}
 is shown as a band. The value of the asy-EoS stiffness parameters has been set to $x$=0 ($L$=61 MeV). 
The same kinematical cuts as in~\figref{rapptspectra} have been applied.}
\end{figure*}

The results presented in Ref.~\cite{Hong:2013yva} have suggested that PMRs are not sensitive to the slope parameter $L$
of the SE at saturation. It was consequently proposed that the kinetic energy (or equivalently transverse momentum) 
dependent PMR, particularly its high energy tail, should be used in order to constraint $L$ from experimental data.
The model employed in that study included a momentum independent isovector
$S$-wave pion potential but the $P$-wave component was completely neglected. Motivated by the upcoming data gathering
campaign of the SAMURAI TPC collaboration~\cite{Shane:2014tsa}, which has among its objectives the experimental 
measurement of pion emission in heavy-ion collisions of impact energies in the neighborhood of the vacuum threshold
for pion production, it is worthwhile to stress once more the importance of including the pion potential with all
its components and realistic energy dependence in transport models that attempt to describe such reactions.
To that end the impact of the pion potential for two values of the impact energy,
$T_{lab}$=250 MeV/nucleon and  $T_{lab}$=400 MeV/nucleon, in mid-central $^{197}$Au+$^{197}$Au collisions
is presented in~\figref{ptspectrapipotdep}. The strength of the isovector $\Delta$(1232) potential and the SE stiffness parameter
have been set to $V_v=\delta$ and $x=0$ respectively. For the higher impact energy case, the FOPI-LAND experimental result~\cite{Reisdorf:2015aa} is also shown
for comparison.

Starting with the $T_{lab}$=400 MeV/nucleon case (right panel of 
~\figref{ptspectrapipotdep}), it is observed that the exclusion of both the $S$- and $P$-wave pion potential
components leads to significant deviations of the model predictions from the experimental data. This is most evident in the 
low-$p_T$ region, where the theoretical model over-predicts the experimental data by a factor of four. In the region surrounding the
average $p_T$ value the discrepancy becomes an under-prediction by a factor of two. Interestingly, the slope of the experimental
data is not reproduced by the no pion potential version of the model even for the highest accessible $p_T$ values. The situation
is considerably improved with the inclusion of the $S$-wave pion potential, particularly in the large $p_T$ region where now both
the value and the slope of the transverse momentum dependent PMR are reasonably close to the experimental data. In the low
$p_T$ region an overestimation of the experimental data by a factor of two persists. This is resolved by the inclusion of the
$P$-wave pion potential (the Batty-1 parametrization in this case), which suggests that the low energy parts of the both $S$- and $P$-wave
potentials are realistic enough. The difference between theoretical and experimental results is however increased in the mid 
and high $p_T$ regions. This result emphasizes once more the importance of the energy dependent part of the pion potential since
for these regions the total pion potential is the result of the subtraction of repulsive $S$-wave and attractive $P$-wave contributions.

These observations are also valid for the case of an impact energy of $T_{lab}$=250 MeV/nucleon (left panel of 
~\figref{ptspectrapipotdep}) with the important difference that the impact of the pion potential on the PMR increases. In the
mid transverse momentum region, the inclusion of the $S$-wave potential enhances the PMR by a factor of two; the subsequent
addition of the $P$-wave potential leads to a decrease by 20-30 $\%$ of this observable. The extraction of a  
trustworthy narrow constraint for the allowed values for $L$ will thus only be possible with the accurate knowledge
and inclusion in the transport model of choice of both components of the pion potential.
The smaller differences between these three cases at higher values for $p_T$ is a direct consequence of the cancellation
of the strengths of the $S$- and $P$-wave components of the pion potential due to the particular assumed momentum dependence
(see~\figref{spwavepot}). 

In contrast to the suggestion put forward or implied by studies of different groups~\cite{Hong:2013yva,Li:2015hfa,Guo:2015tra},
the results presented above indicate that it may be worthwhile, when attempting to extract the slope parameter $L$ from pion observables, 
to apply a kinetic energy (or transverse momentum) cut to pion spectra, including only events below a certain maximum value.
For the case of kinetic energy spectra a conservative value for this upper limit must not be significantly larger than the kinetic energy
of pions in pionic atoms, limiting it to values as low as 20-30 MeV. This ensures that uncertainties in the energy dependent part of the 
pion potential are largely removed. 
Uncertainties in the density dependence above the saturation point of the potential survive and should be removed by means other than 
kinematical cuts applied on spectra, since it is this density region that provides the interesting physics signal one seeks to isolate.
A possible solution to this issue will require identification of heavy-ion observables that present
an enhanced sensitivity to the strength of the pion potential and a suppressed one to the quantities of interest. In
this context, heavy-ion experimental data for nuclei with small isospin asymmetry may prove valuable. 


\begin{figure*}
 \includegraphics[width=26.0pc]{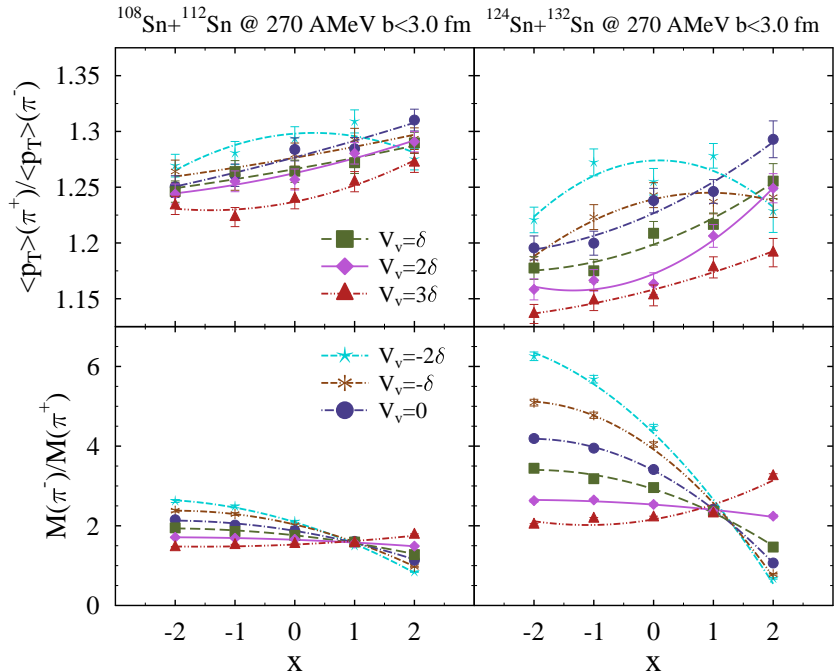}
\caption{\figlab{samuraitpcpred}(Color online) Theoretical predictions for pion multiplicity (lower panels) and
average transverse momentum (upper panels) ratios in $^{108}$Sn+$^{112}$Sn (left panels)
and $^{124}$Sn+$^{132}$Sn (right panels) central collisions at an impact energy of 270 MeV/nucleon as a function
of the stiffness parameter $x$ and for the various choices of the isovector $\Delta$(1232) isobar discussed
in the text. No kinematical cuts have been applied on spectra.}
\end{figure*}

The present section is concluded with predictions relevant for extracting the density dependence of the SE from
forthcoming experimental data of the SAMURAI TPC Collaboration. Measurements of pion production 
in collisions of various combinations of Sn isotopes for a projectile laboratory impact energy 
of $T_{lab}$=270 MeV/nucleon $^{108}$Sn+$^{112}$Sn have been performed recently~\cite{Mu:2016aa}
and experimental results for the multiplicity and average $p_T$ ratios will presumably become available 
during the next couple of years. Of these, predictions for those reactions
involving nuclei that lead to fireballs with the lowest and highest possible isospin asymmetry,
 $^{108}$Sn+$^{112}$Sn and $^{132}$Sn+$^{124}$Sn (projectile+target) respectively~\cite{Mu:2016aa}, will be presented.
The nuclei initialization part of the model has been tuned such as to reproduce, on average, 
the experimentally measured rms of each of the mentioned isotopes~\cite{Angeli:2013aa} and generate 
initial density profiles that reproduce a two-parameter Fermi distribution with a diffuseness 
parameter a=0.55 fm~\cite{DeJager:1987qc}.

The predictions for the two mentioned reactions for central collisions (b$<$3.0 fm) and without any kinematical cuts 
are presented in~\figref{samuraitpcpred}. The global energy conservation version of the model, 
including the effective $S$-wave ChPT inspired and Batty-1 $P$-wave pion potentials, has been employed.
The difference in sensitivity to both the isovector nucleon and $\Delta$(1232) potentials between the 
low ($\beta$=0.09) and high ($\beta$=0.22) isospin asymmetry cases  is clearly observed for both the PMR
and PAPTR, being significantly more pronounced for the latter isospin asymmetry choice. Compared to $^{197}$Au+$^{197}$Au collisions
at 400 MeV/nucleon impact energy (see~\figref{avptratdelpotdep}) the increase, if any, of the sensitivity of the PMR 
to the SE stiffness is a function of the strength of the isovector $\Delta$(1232) potential, being more pronounced
for negative values of $V_v$. For the values of the parameters $x$ and $V_v$ favored by
the existing FOPI experimental data the increase of the sensitivity is in the range of 10-20$\%$.
A comparison of PMR with a $\Delta$ resonance model predictions that assumes only first chance inelastic collisions
~\cite{Stock:1985xe}, $M(\pi^-)/M(\pi^+)=(5N^2+NZ)/(5Z^2+NZ)\approx(N/Z)^2$, reveals that these transport model 
calculations overpredict that result with similar factors for both reactions, similarly to other models~\cite{Xiao:2008vm}.
The magnitude of these differences is found to depend on both the SE stiffness and 
the strength of the isovector $\Delta$(1232) potential.

\section{Summary and Conclusions}
\seclab{sumcon}
A QMD type transport model applicable to heavy-ion collisions with impact energies of a few hundred MeV/nucleon that
allows the conservation of the total energy of the system during such reactions has been further extended by including
the effects of the $S$- and $P$-wave components of the pion optical potential. This allows theoretical
computations of observables related to the final momenta of the emitted pions. Of these, the final average transverse 
momenta of charged pions and their ratio (PAPTR) have been studied in detail. The main result of this study is 
the proof of feasibility of using this observable in conjunction with the charged pion multiplicity ratio (PMR) to extract constraints,
from experimental data, for both the stiffness of the symmetry energy and the strength of the isovector component of the
$\Delta$(1232) potential in nuclear matter.

It has been shown that the energy conservation scenarios introduced in a previous study have a significant impact on the
ratios of average transverse momenta. Specifically, in the case when the effect of potential energies is not taken into
account in the energy conservation constraint appearing in the collision term of the transport equations, the theoretical value of PAPTR is systematically overestimating the experimental one by about 20$\%$.
Within the local energy conservation scenario (LEC) the value of PAPTR underestimates the experimental data by a fraction
in the range of  5-10$\%$. Finally, requiring the conservation of the total energy of the system (GEC scenario) leads to
an increase of the value of PAPTR such that the experimental data are once again overestimated by an amount in the
range of 10-15$\%$.

The $S$-wave component of the pion potential that has been employed in this study takes into account theoretical results concerning 
chiral symmetry restoration and pion-nucleon scattering in dense nuclear matter that have been previously validated
by the experimental study of properties of pionic atoms. An energy dependence of the potential has been
incorporated by using as guidance empirical results about the behavior of the vacuum isoscalar and isovector scattering 
pion-nucleon amplitudes away from threshold and requiring an agreement with experimental data on pion-nucleus
scattering up to pion kinetic energies of 300 MeV.

The $P$-wave pion optical potentials used in this work have their origin in pionic atom studies. Their extrapolation
to higher momenta must however be handled with care, given the strong energy dependence of the $\Delta$(1232) width. To
this end the energy dependence of a theoretical model for the optical pion potential derived within the
framework of the delta-hole approximation has been mirrored . To be on par with theoretical models describing pionic atom properties, also the density
gradient terms of the potential have been implemented in the transport model. They have been shown to have a non-negligible
impact on observables in the context of constraining the strength of the isovector $\Delta(1232)$ potential.

The total impact of the $S$- and $P$-wave components of the pion potential on pion multiplicity ratio is 
found to be moderate. However, by neglecting their contribution, uncertainties that can amount to as 
much as 50 MeV on the extracted value for the symmetry energy slope at saturation $L$ could be expected,
at fixed isovector $\Delta$(1232) potential strength. The presented model for the pion optical potential
leads to a good description of experimental values for the average transverse momenta, at 5$\%$ level,
in mid-central $^{197}$Au+$^{197}$Au at an impact energy of 400 MeV/nucleons. The description of the experimental 
value of the ratio of average transverse momenta of charged pions is of even better quality for certain choices of the $P$-wave optical
potential parametrization.

The impact of the strength of the isovector $\Delta$(1232) potential on the pion multiplicity and average transverse momentum
ratios is investigated, confirming the conclusions of a previous study for the PMR and evidencing a similar behavior for the PAPTR. 
Using available FOPI experimental data for these observables, it is shown that constraints for both the slope $L$ of the symmetry energy
at saturation and the strength of the isovector $\Delta$(1232) potential can be extracted from a two dimensional $\chi^2$ fit. 
The inclusion of the $S$- and $P$-wave pion potentials, particularly the former one, is found to be crucial 
for a simultaneous description of FOPI experimental data for the pion multiplicity and average $p_T$ ratios
to be possible for realistic values of the stiffness of the symmetry energy and strength of the isovector $\Delta$(1232)
potential.
The obtained values for these parameters are however rather imprecise due to rather large uncertainties that 
affect the experimental data (multiplicity ratios) and inaccurate knowledge of the pion optical potential for the entire
density and energy range probed in heavy-ion collisions.

The presently available experimental data favor a value of the slope parameter $L$ larger than 50 MeV, at 1$\sigma$ confidence level, 
implying the claim that the symmetry energy is not soft. The allowed upper limit is however a very stiff one, exceeding the value 
$L$=150 MeV for any choice of the $P$-wave pion potential. Additionally, the allowed
value for the isovector $\Delta$(1232) potential strength is, on average, somewhat weaker, by about 25$\%$, than the usual
choice employed in transport models. For the favored values for these two parameters, the transport model
allows a good description of available FOPI experimental rapidity and transverse momentum multiplicity spectra 
of charged pions in mid-central $^{197}$Au+$^{197}$Au collisions at an impact energy of 400 MeV/nucleon.

It has also been shown that by increasing the precision of the experimental measurements
the uncertainties of the extracted constraints for the stiffness of the symmetry energy can be significantly reduced. 
The situation can be further improved by studying experimentally systems with a higher isospin asymmetry at 
impact energies closer to the vacuum pion production threshold than previously accomplished.
For such a program to be successful the knowledge of the pion potential at densities higher than
probed in pionic atom experiments and its energy dependence will have to be advanced. This is necessary also because the impact
of the pion optical potential increases as the collision energy is decreased. In this context, experimental measurements of 
reactions involving isospin symmetric heavy-nuclei and restriction of particle spectra to low kinetic energy pions may prove helpful.

\begin{acknowledgments}
The research of M.D.C. has been financially supported by the Romanian Ministry of Education and
Research through Contract No. PN 16420101/2016. The author would like thank Prof. Dr. W. Reisdorf for providing
yet unpublished FOPI experimental data.
The assistance of the DFCTI department of IFIN-HH with the maintenance of the computing
cluster on which simulations were performed is gratefully acknowledged.
\end{acknowledgments}

\bibliography{references}

\end{document}